\documentclass[iop,revtex4,letterpaper]{emulateapj}
\usepackage{natbib}
\usepackage[colorlinks=true,urlcolor=blue,citecolor=blue,linkcolor=blue,breaklinks]{hyperref}
\usepackage{amsmath}
\usepackage{pdflscape}

\begin{document}

\newcommand{\nnhp}{N$_2$H$^+$}
\newcommand{\nthp}{N$_2$H$^+$}
\newcommand{\hcop}{HCO$^+$}
\newcommand{\hcn}{HCN}
\newcommand{\hnc}{HNC}
\newcommand{\hctn}{HC$_3$N}
\newcommand{\cth}{C$_2$H}
\newcommand{\kms}{km$\,$s$^{-1}$}
\newcommand{\Kkms}{K\,km$\,$s$^{-1}$}
\newcommand{\Kkmspcsq}{\Kkms\,pc$^2$}
\newcommand{\lhcn}{$L_{\rm{HCN}}$}
\newcommand{\lhcop}{$L_{\rm{HCO^+}}$}
\newcommand{\lhnc}{$L_{\rm{HNC}}$}
\newcommand{\lnthp}{$L_{\rm{N_2H^+}}$}

\newcommand{\ltcs}{$L_{\rm{ ^{13}CS}}$}
\newcommand{\lhfoa}{$L_{\rm{H41\alpha}}$}
\newcommand{\lchtcn}{$L_{\rm{CH_3CN}}$}
\newcommand{\lhctn}{$L_{\rm{HC_3N}}$}
\newcommand{\ltctfs}{$L_{\rm{^{13}C^{34}S}}$}
\newcommand{\lhctccn}{$L_{\rm{HC^{13}CCN}}$}
\newcommand{\lhncofzf}{$L_{\rm{HNCO 4_{0,4}}}$}
\newcommand{\lcth}{$L_{\rm{C_2H}}$}
\newcommand{\lhntc}{$L_{\rm{HN^{13}C}}$}
\newcommand{\lsio}{$L_{\rm{SiO}}$}
\newcommand{\lhtcop}{$L_{\rm{H^{13}CO^+}}$}

\newcommand{\fhcn}{$F_{\rm{HCN}}$}
\newcommand{\fhcop}{$F_{\rm{HCO^+}}$}
\newcommand{\fhnc}{$F_{\rm{HNC}}$}
\newcommand{\fnthp}{$F_{\rm{N_2H^+}}$}
\newcommand{\ftcs}{$F_{\rm{ ^{13}CS}}$}
\newcommand{\fhfoa}{$F_{\rm{H41\alpha}}$}
\newcommand{\fchtcn}{$F_{\rm{CH_3CN}}$}
\newcommand{\fhctn}{$F_{\rm{HC_3N}}$}
\newcommand{\ftctfs}{$F_{\rm{^{13}C^{34}S}}$}
\newcommand{\fhctccn}{$F_{\rm{HC^{13}CCN}}$}
\newcommand{\fhncofzf}{$F_{\rm{HNCO 4_{0,4}}}$}
\newcommand{\fcth}{$F_{\rm{C_2H}}$}
\newcommand{\fhntc}{$F_{\rm{HN^{13}C}}$}
\newcommand{\fsio}{$F_{\rm{SiO}}$}
\newcommand{\fhtcop}{$F_{\rm{H^{13}CO^+}}$}

\newcommand{\lhcneg}{$L_{\rm{HCN,eg}}$}
\newcommand{\lhcnmask}{$L_{\rm{HCN,mask}}$}
\newcommand{\lhcnoz}{$L_{\rm{HCN}(1-0)}$}
\newcommand{\lcott}{$L_{\rm{CO}(3-2)}$}
\newcommand{\lcooz}{$L_{\rm{CO}(1-0)}$}
\newcommand{\hcnoz}{HCN(1--0)}

\newcommand{\cooz}{\mbox{CO(1--0)}}
\newcommand{\nthpoz}{\mbox{N$_2$H$^+$(1--0)}}
\newcommand{\hcopoz}{\mbox{HCO$^+$(1--0)}}
\newcommand{\hncoz}{\mbox{HNC(1--0)}}
\newcommand{\hctntn}{\mbox{HC$_3$N(10--9)}}
\newcommand{\cthoz}{\mbox{C$_2$H(1--0)}}

\newcommand{\lsun}{$L_\odot$}
\newcommand{\lir}{$L_{\rm{IR}}$}
\newcommand{\lfir}{$L_{\rm{FIR}}$}
\newcommand{\liriras}{$L_{\rm{IR}}$}
\newcommand{\firiras}{$F_{\rm{IR}}$}
\newcommand{\lirhersch}{$L_{\rm{IR,Hersch}}$}
\newcommand{\lmol}{$L_{\rm{molecule}}$}
\newcommand{\hncofzf}{HNCO(4$_{0,4}$ -- 3$_{0,3}$)} 
\newcommand{\gsr}{Gao--Solomon relation}
\newcommand{\mvir}{$M_{\rm{Vir}}$}
\newcommand{\siggas}{$\Sigma_{\rm{gas}}$}
\newcommand{\sigsfr}{$\Sigma_{\rm{SFR}}$}
\newcommand{\mdense}{$M_{\rm{dense}}$}
\newcommand{\funit}{\liriras/\mdense}
\def\pasa{PASA}

\title{Linking Dense Gas from the Milky Way to External Galaxies} 

\author{Ian W. Stephens\altaffilmark{1}, James M. Jackson\altaffilmark{1}, J. Scott Whitaker\altaffilmark{2}, Yanett Contreras\altaffilmark{3},
Andr\'es E. Guzm\'an\altaffilmark{4}, Patricio Sanhueza\altaffilmark{2,5}, Jonathan B. Foster\altaffilmark{1,6}, Jill M. Rathborne\altaffilmark{3}}
\altaffiltext{1}{\itshape Institute for Astrophysical Research, Boston University, Boston, MA 02215, USA
ianws@bu.edu}
\altaffiltext{2}{\itshape Physics Department, Boston University, Boston, MA 02215, USA}
\altaffiltext{3}{\itshape Australia Telescope National Facility, CSIRO Astronomy and Space Science, Epping, NSW, Australia}
\altaffiltext{4}{\itshape Departamento de Astronom\'ia, Universidad de Chile, Camino el Observatorio 1515, Las Condes, Santiago, Chile}
\altaffiltext{5}{\itshape National Astronomical Observatory of Japan, 2-21-1 Osawa, Mitaka, Tokyo 181-8588, Japan}
\altaffiltext{6}{\itshape Yale Center for Astronomy and Astrophysics, New Haven, CT 06520, USA}

\interfootnotelinepenalty=10000

\begin{abstract}
In a survey of 65 galaxies, \citet{GaoSolomon2004} found a tight linear relation between the infrared luminosity (\lir, a proxy for the star formation rate) and the \hcnoz\ luminosity (\lhcn). \citet{Wu2005,Wu2010} found that this relation extends from these galaxies to the much less luminous Galactic molecular high-mass star-forming clumps ($\sim$1~pc scales), and posited that there exists a characteristic ratio \lir /\lhcn\ for high-mass star-forming clumps. The \gsr\ for galaxies could then be explained as a summation of large numbers of high-mass star-forming clumps, resulting in the same \lir /\lhcn\ ratio for galaxies. We test this explanation and other possible origins of the \gsr\ using high-density tracers (including \hcnoz, \nthpoz, \hcopoz, \hncoz, \hctntn, and \cthoz) for $\sim$300 Galactic clumps from the Millimetre Astronomy Legacy Team 90 GHz (MALT90) survey. 
The MALT90 data show that the \gsr\ in galaxies cannot be satisfactorily explained by the blending of large numbers of high-mass clumps in the telescope beam. Not only do the clumps have a large scatter in the \lir/\lhcn\ ratio, but also far too many high-mass clumps are required to account for the Galactic IR and HCN luminosities. We suggest that the scatter in the \lir/\lhcn\ ratio converges to the scatter of the \gsr\ at some size-scale $\gtrsim$1~kpc.
We suggest that the \gsr\ could instead result from of a universal large-scale star formation efficiency, initial mass function, core mass function, and clump mass function.

\end{abstract}
\subjectheadings{stars: formation -- stars: massive -- galaxies: star formation -- ISM: molecules -- ISM: clouds}

\maketitle

\section{Introduction} Ê\label{sec:intro}
The Kennicutt--Schmidt law \citep{Schmidt1959,Schmidt1963,Kennicutt1998a,Kennicutt1998b} describes an empirical relation between the surface density of star formation, \sigsfr, and the surface density of gas, \siggas, in the form \sigsfr~$\propto$~(\siggas)$^N$. By tracing gas in normal spiral and starburst galaxies with H$\alpha$, \ion{H}{1}, and CO(1--0), $N$ (commonly referred to as the Schmidt index) was found to be $1.4\pm0.15$. Stars, however, form in dense clumps (size-scales of $\sim$1~pc) in giant molecular clouds (GMCs). While the gas mass of GMCs is best traced by CO, the dense clumps are better traced by molecules with higher dipole moments, such as HCN and HCO$^+$, since these molecules are collisionally excited into emission only at higher densities. \citet[][henceforth, GS04]{GaoSolomon2004} investigated the Kennicutt--Schmidt law using the high-density tracer HCN(1--0) in normal galaxies, luminous infrared galaxies (LIRGs), and ultraluminous infrared galaxies (ULIRGs). Specifically, GS04 summed the fluxes across entire galaxies to calculate the \hcnoz\ luminosity (\lhcn) and the infrared (IR) luminosity (\lir, as derived from the Infrared Astronomical Satellite, i.e., IRAS). The luminosities \lir\ and \lhcnoz\ have a tighter power-law correlation than \lir\ and \lcooz. Remarkably, the relations have significantly different power-laws; while \lir~$\propto$~\lcooz$^{1.4}$ (i.e., follows the Schmidt index $N=1.4$), \lir~$\propto$~\lhcnoz$^{1.00\pm0.05}$. Specifically, GS04 found the empirical power-law log\,\liriras\,=\,1.00\,log \lhcn\,+\,2.9, where \liriras\ is the IRAS IR luminosity in units of $L_\sun$ and \lhcn\ is the HCN(1--0) luminosity in units of \Kkmspcsq. This relation will henceforth be called the \gsr. \citet{Narayanan2005} investigated the same relation with CO(3--2) rather than HCN(1--0) and found a roughly consistent power-law \lcott~$\propto$~\lir$^{0.92\pm0.07}$.  

\citet[][henceforth, Wu05]{Wu2005} observed \hcnoz\ for $\sim$50 Galactic star-forming clumps, most of which are forming high-mass stars. When they compared these Galactic clumps ($\sim$1~pc scale) to the galaxies of GS04, they found that the \gsr\ extends over several orders of magnitude to these Galactic clumps. A similar result was found for dense gas within molecular clouds \citep{Lada2012}. A preliminary study based on the Millimetre Astronomy Legacy Team 90 GHz (MALT90) survey confirmed the Wu05 results \citep{Jackson2013}. The interpretation set forth by Wu05 was that the extragalactic luminosities \liriras\ and \lhcnoz\ could be explained as a summation of the luminosities of all high-mass star-forming clumps within a galaxy.  Wu05 also suggested that the relation no longer holds for clumps with \liriras\ below $10^{4.5} L_\sun$ -- approximately the luminosity of an ultracompact \ion{H}{2} region. They suggested that clumps below this luminosity may not contain high-mass stars and therefore do not sample the complete initial mass function (IMF). 

\citet{KrumholzThompson2007} suggested that, given a turbulent medium, the relation between \liriras\ and a molecular line's luminosity depends on the transition's critical density and the median density of a galaxy. For the relation \liriras~$\propto$~\lmol$^\alpha$, lines with critical densities lower than a galaxy's median density of star-forming clouds, such as \cooz, will have $\alpha$ equal to the Schmidt index $N$. In contrast, lines with critical densities higher than this median density, such as \hcnoz, will have $\alpha < N$. \citet{Narayanan2008} took a different approach to analyze the relation via 3D hydrodynamic simulations and found similar results as \citet{KrumholzThompson2007}. They also predicted $\alpha$ for higher $J$ rotational transitions of CO and HCN. HCN(3--2), for example, is expected to have $\alpha \approx 0.7$. HCN(3--2) observations of galaxies follow $\alpha \sim 0.8$ \citep{Gracia2008, Bussmann2008, Juneau2009}. Using observations of Galactic star-forming clumps, \citet[][henceforth, Wu10]{Wu2010} found that HCN(3--2) has $\alpha=0.88 \pm 0.06$, significantly lower than that for \hcnoz\ observations ($\alpha = 1.07 \pm 0.06$), but still higher than that predicted for galaxies in \citet{Narayanan2008}.

The physical basis for the \gsr\ is still under debate. Wu05 and Wu10 suggested that if the clump samples the high-mass regime of the stellar IMF, the clump will have a characteristic value of \lir/\lhcn\ (a proxy measurement for \funit, where \mdense\ is the dense gas mass). Summing large numbers of clumps with this characteristic value will preserve a linear relation from the clump-scale to the galaxy-scale. \citet{KrumholzThompson2007}, however, suggested that such a relation will occur over a large continuous distribution of interstellar medium structures given a universal star formation efficiency. A better understanding of the physical basis for the \gsr\ can be achieved by increasing the Galactic sample size of Wu05 and Wu10 and by observing other molecules that also trace high densities. The MALT90 survey \citep{Foster2011,Foster2013,Jackson2013} targeted over 3000 star-forming clumps identified in the ATLASGAL 870~$\mu$m continuum survey \citep{Schuller2009}, simultaneously observing 16 lines near 90~GHz.  Approximately 85\% of these clumps surveyed have masses above 200~$M_\odot$ (Y. Contreras in preparation), and clumps above 200~$M_\odot$ are likely to form high-mass stars \citep[e.g.,][]{Jackson2013}. Thus, the MALT90 survey can test the relation between \lir\ and molecular line luminosity with a much larger sample of high-mass star-forming clumps than the Galactic sample of $\sim$50 clumps studied by Wu05. 



In this paper, we investigate the \gsr\ using $\sim$300 clumps that were covered by the MALT90 survey. We investigate the relation in a similar manner as GS04, Wu05, and Wu10, which compares the IR luminosity calculated from IRAS fluxes with the molecular line luminosity. In Section~\ref{sec:data} we discuss the data used and the calculations of IR and molecular line luminosities. In Section~\ref{gsrHCN}, we discuss the \gsr, concentrating only on the \hcnoz\ molecular line. Specifically, we find that the \gsr\ does not seem to be well explained by a summation of high-mass clumps, and we discuss other possible origins of IR and \hcnoz\ emission and the \gsr\ itself. In Section~\ref{gsrall} we discuss the \gsr\ with respect to the other dense gas tracers that were covered by the MALT90 survey. In Section~\ref{discussion} we discuss the implications of our analysis, and in Section~\ref{summary} we summarize the results.

\section{Data and Methodology}\label{sec:data}

\begin{deluxetable}{lc@{}c}
\tablecolumns{3}
\tabletypesize{\small}
\tablewidth{0pt}
\tablecaption{Spectral Lines Covered by the MALT90 Survey \label{tab:m90}}
\tablehead{\colhead{Species} & \colhead{Main transition} & \colhead{Frequency (GHz)}
}
\startdata
\nthp & \emph{J} = 1 -- 0 & 93.173772 \\
 $^{13}$CS &  \emph{J} = 2 -- 1 &  92.494303 \\
H & $41\alpha$ & $92.034475$\tablenotemark{*} \\
CH$_3$CN & $J_K$ = 5$_1$ -- 4$_1$ &  91.985313\\ 
HC$_3$N & $J$ = 10 -- 9 & 90.979020 \\  
$^{13}$C$^{34}$S & $J$ = 2 -- 1 & 90.926036\tablenotemark{*} \\
HNC & \emph{J} = 1 -- 0 & 90.663572 \\
HC$^{13}$CCN & $J$ = 10 -- 9, $F$ = 9 -- 8 & 90.593059\tablenotemark{*} \\
HCO$^+$ & \emph{J} = 1 -- 0 & 89.188526 \\
HCN & $J$ = 1 -- 0 & 88.631847 \\
HNCO & $J_{K_a,K_b}$ = 4$_{1,3}$ -- 3$_{1,2}$ & 88.239027\tablenotemark{*}  \\
HNCO & $J_{K_a,K_b}$ = 4$_{0,4}$ -- 3$_{0,3}$ & 87.925238 \\
C$_2$H & $N$ = 1 -- 0, $J$ = $\frac{3}{2} - \frac{1}{2}$, & 87.316925 \\
& $F$ = 2 -- 1 & \\
HN$^{13}$C & $J$ = 1 -- 0 & 87.090859 \\
SiO & $J$ = 2 -- 1 & 86.847010 \\
H$^{13}$CO$^+$ & $J$ = 1 -- 0 & 86.754330
\enddata
\tablenotetext{*}{These spectral lines that were not used in this study.} 
\end{deluxetable}

MALT90 mapped 16 lines for 3246 clumps, primarily high-mass star-forming clumps that are $>$200~$M_\odot$ \citep{Jackson2013}, as identified from the ATLASGAL survey. Many of these clumps had two velocity components separated significantly in velocity, indicating two unrelated clumps along the line of sight. The MALT90 catalog (J. Rathborne in preparation) considers sources that have two velocity components separated by more than 15~\kms\ as two separate clumps, which brings the MALT90 catalog to a total of 3566 clumps (J. Rathborne in preparation). From henceforth, we will refer to one of these 3566 clumps as a ``MALT90 clump." In Table \ref{tab:main}, these velocity-separated clumps are indicated using ``\_A" and ``\_B" suffixes. Clumps with the suffix ``\_S" are sources that do not have multiple velocity components separated by 15~\kms.

The lines observed with the MALT90 survey are given in Table~\ref{tab:m90}. The antenna temperature, $T_A^*$, for all fluxes were converted to main beam temperature via $T_{\rm{mb}} = T_A^*/\eta_{\rm{mb}}$ where the main beam efficiency $\eta_{\rm{mb}}$ is taken to be 0.5 \citep{Ladd2005}. For brevity in the rest of the paper, we will not show the transitions for the molecules, and the transitions indicated in Table \ref{tab:m90} will be referred to by the name of the molecule unless otherwise specified. Maps of the MALT90 survey are all lightly smoothed so that the full-width at half-maximum (FWHM) beam size is $\theta_{\rm{beam}}$~=~37$\farcs$8 at all frequencies.

In this paper we will define high-mass clumps to have masses $>$200~$M_\odot$, which are likely to form stars with masses $>$8~$M_\odot$ \citep{Jackson2013}, and define any clump $<$200~$M_\odot$ as a low-mass star-forming clump. 


\subsection{Calculation of Infrared Luminosity and Molecular Line Luminosity for MALT90 Clumps}\label{sec:calcs}

In order to compare luminosities derived from IRAS (\liriras) to molecular line luminosities from MALT90 (\lmol), we first matched the MALT90 clumps to the IRAS Point Source Catalog v2.1 (PSC). Although the MALT90 survey covered 3566 clumps, not all of these clumps have a corresponding IRAS source since compact sources may be extremely beam diluted and the wavelengths probed by IRAS are too short to probe the coldest clumps. In matching MALT90 clumps to IRAS sources, we required that the position of the MALT90 clump is within 0\fdg005 (18$\arcsec$; approximately half the size of the MALT90 beam) of the IRAS point source position. This cutoff was chosen since larger separations often include IRAS sources that are not associated with the MALT90 clump, and smaller separations remove sources that are associated with each other. 

Following GS04, Wu05, and Wu10, we used the standard formula to calculate \lir\ from IRAS fluxes \citep{Sanders1996} 
 
\begin{equation}
L_{\rm{IR}}  = 0.56D^2 (13.48f_{12} + 5.16f_{25} + 2.58f_{60} + f_{100}), \label{eqliriras}
\end{equation}
where \liriras\ is in $L_\sun$, distance $D$ is in kpc, and $f_x$ is the flux of the $x~\mu$m IRAS band in Jy. This equation assumes a single temperature dust emissivity model. For dust temperatures in the range of 25--65~K, the luminosity should be accurate within 5\%  \citep{Sanders1996}. Kinematic distances for \liriras\ (and \lhcn, see below) were calculated based on the clump velocity and solving for the near/far ambiguity, which will be presented in a future paper (J. Whitaker et al. in preparation).


In order to calculate line luminosities of MALT90 clumps with the same methodology used in Wu05 and Wu10, for each line we fit each IRAS-matched MALT90 integrated intensity (moment 0) map with an elliptical Gaussian. We associated 405 MALT90 clumps with an IRAS source; a large fraction of the MALT90 clumps are cold \citep{Guzman2015}, beam diluted with the IRAS, or distant and thus are undetected with IRAS. A MALT90 map is typically $4\arcmin \times 4\arcmin$ ($27\times27$ pixels of size $9\arcsec\times9\arcsec$), but the outer 0$\farcm$5 of the maps are noisier since the edges of the map have significantly less integration time due to the Mopra on-the-fly mapping procedure. Therefore, our fitting routine only considers the inner 75\% of the pixels in each dimension (i.e., $\sim3\arcmin \times 3\arcmin$) for each MALT90 map. The fitting routine also required that there be at least 5 pixels across the entire map with a signal-to-noise greater than 6. After all moment 0 maps were fit with an elliptical Gaussian, we manually inspected each fit overlaid on the moment 0 maps to judge whether the elliptical Gaussian provided a reasonable fit. Fits were removed if: (1) the map was significantly noisy, typically due to observations in poor weather that resulted in bad baselines; (2) a large portion ($\gtrsim$20\%) of the elliptical gaussian's FWHM extended beyond the map boundaries; (3) multiple peaks in the MALT90 map led to poor elliptical Gaussian fits; (4) the centroid of the elliptical Gaussian fit was $\gtrsim$2$\arcmin$ from the clump and thus not likely to be associated with the IRAS source; (5) a MALT90 clump corresponded with two IRAS sources; or (6) the emission within the MALT90 map was too smooth and extended to allow for a proper elliptical Gaussian fit. Of the 405 IRAS-matched MALT90 clumps, 282 clumps had a valid flux for at least one line. Twelve of the sixteen spectral lines are presented in this paper. \mbox{HNCO(4$_{1,3}$ -- 3$_{1,2}$)}, \mbox{HC$^{13}$CCN($J$~=~10--9, $F$~=~9--8)}, and \mbox{$^{13}$C$^{34}$S(2--1)} had no valid elliptical Gaussian fits and H41$\alpha$ is not a molecular line and had only three valid elliptical Gaussian fits.



Given the elliptical Gaussian fit of a map, we can calculate the integrated intensity in a similar manner as GS04. We follow the equation given in Wu05 and Wu10

\begin{eqnarray} 
L_{\rm{molecule}} = 23.5 \times 10^{-6} D^2  \frac{\pi \theta_s^2}{4\,\rm{ln}\,2} \frac{\theta_s^2 + \theta_{\rm{beam}}^2}{\theta_s^2} \int T_{\rm{mb}}^* \,dv  \nonumber \\ 
= 23.5 \times 10^{-6} D^2  \frac{\pi \theta_{\rm{fit}}^2}{4\,\rm{ln}\,2}   \int T_{\rm{mb}}^* \,dv,~~~~~~~~~~~~~ \label{eqlmol}
\end{eqnarray}
where \lmol\ is in units of \Kkmspcsq, $D$ is in kpc, $\theta_s$ and $\theta_{\rm{beam}}$ is the actual (deconvolved) angular size of the source and the angular size of the beam, respectively, in arcseconds, and $\int T_{\rm{mb}}^* \,dv$ is the peak velocity-integrated intensity in units of \Kkms. $\theta_s^2 + \theta_{\rm{beam}}^2$ is equal to the Gaussian fit size, which is approximated to be $\theta_{\rm{fit}} = \sqrt{\theta_{\rm{maj}} \times \theta_{\rm{min}}}$, where $\theta_{\rm{maj}}$ and $\theta_{\rm{min}}$ are the FWHM sizes of the major and minor axis of the elliptical Gaussian fit. 

The results of the elliptical Gaussian fits are given in Table~\ref{tab:main}. In this table, we only show fluxes for the MALT90 clumps rather than luminosities since certain clumps do not have valid kinematic distances. The fluxes reported in this paper are simply the luminosities divided by the square of the distance in kpc$^{2}$, i.e., 

\begin{equation}
F_{\rm{IR}} = 0.56 (13.48f_{12} + 5.16f_{25} + 2.58f_{60} + f_{100}), \label{eqfiriras}
\end{equation}

\begin{equation}
F_{\rm{molecule}} = 23.5 \times 10^{-6} \frac{\pi \theta_{\rm{fit}}^2}{4\,\rm{ln}\,2}   \int T_{\rm{mb}}^* \,dv  , \label{eqfmol}
\end{equation}
where variables in these two equations are in the same units as Equations \ref{eqliriras} and \ref{eqlmol}. The units for $F_{\rm{IR}}$ and $F_{\rm{molecule}}$ are therefore $L_\odot \, \rm{kpc}^{-2}$ and \Kkmspcsq $\,$kpc$^{-2}$, respectively. The latter has both pc and kpc in the unit; we chose to keep the units in this format for easy conversion to luminosity by simply multiplying the fluxes by the square of the kpc distance $D$ given in Column 3 of Table~\ref{tab:main}. $F_{\rm{molecule}}$ can be converted from \Kkmspcsq $\,$kpc$^{-2}$ to \Kkms\ by multiplying by 10$^{-6}$.

Of the 282 clumps that had a valid flux for at least one line, 209 clumps also had a valid kinematic distance and thus have a luminosity for at least one line. Most of the clumps without a valid kinematic distance are those within 10$^\circ$ of the Galactic Center; kinematic distances within this longitude range cannot be calculated with high accuracy since a slight change in velocity can dramatically change a kinematic distance. Of these 209 clumps, 204 clumps had only one velocity component (i.e., has a suffix ``\_S" in Table \ref{tab:main}). We only consider the molecular line luminosities for these 204 clumps throughout the paper.

\subsection{Data from GS04 and Wu10}
For the figures in this paper, we include data from GS04 and Wu10. From Wu10, we do not include the clump DR21S since the value reported in their table for \lhcn\ is obviously errant (the value is too low for the reported flux, size, and distance of the source). 

\begin{figure}[ht!]
\begin{center}
\includegraphics[width=1\columnwidth]{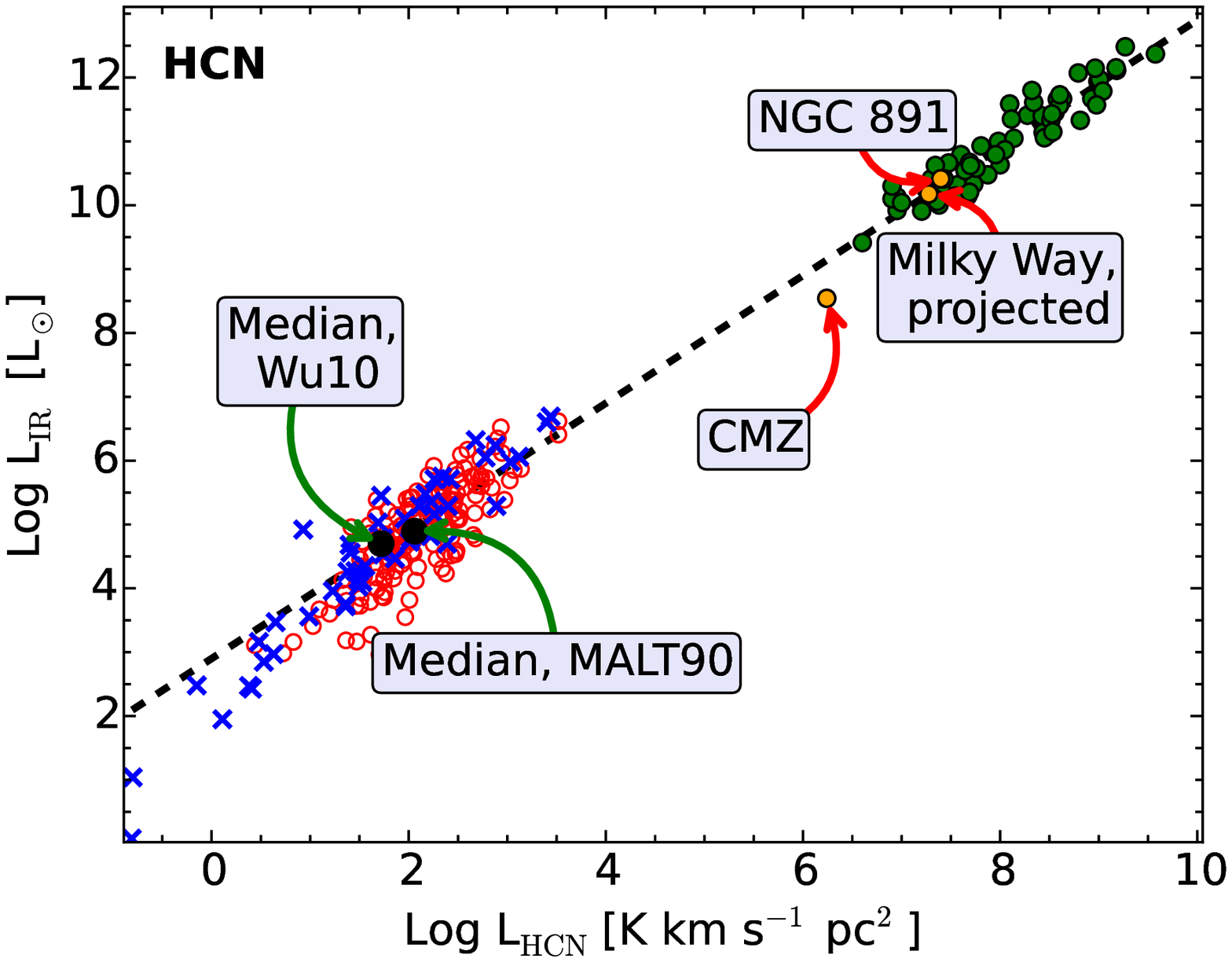}
\end{center}
\caption{The \gsr\ (\liriras\ versus \lhcnoz) shown for galaxies and Galactic clumps. Green circles are galaxies from GS04, blue crosses are Galactic clumps from Wu10, and red circles are MALT90 Galactic clumps presented in this paper. The black dashed line shows the original \gsr\ fit of log \liriras\ = 1.00 log \lhcn\ + 2.9, i.e., it is not fitting any of the Galactic sources. The relation extends from galaxies to Galactic clumps. The median values for the Wu10 and MALT90 clumps are indicated, as well as NGC~891. We show the projected location of the Milky Way assuming \liriras\ = $1.5 \times\ 10^{10}$~$L_\odot$ \citep{Cox1986} and that the Milky Way follows the \gsr. We also show the location of the Central Molecular Zone (CMZ) with intensities integrated over the solid angle subtended by the Mopra CMZ 3\,mm Band Survey discussed in Section~\ref{largecmz}.
}
\label{gaosolomon_IRASimfit_HCNonly} 
\end{figure}

\section{The IR and HCN luminosity correlation for Galactic Clumps}\label{gsrHCN}
Of the 204 IRAS-matched MALT90 clumps with valid luminosities for at least one line (see Section \ref{sec:calcs}), 160 had valid HCN luminosities. In Figure~\ref{gaosolomon_IRASimfit_HCNonly} we show the \gsr\ along with the clumps from Wu10 and MALT90 (this study).  
Like Wu05, Wu10, and \citet{Jackson2013}, we find that the fit to the \gsr\ extends from the higher luminosities found in galaxies to the lower luminosities found in Galactic clumps.  

\subsection{The Clump Discrepancy}\label{clumpsum}
Based on any standard IMF \citep[e.g.,][]{Salpeter1955,Kroupa2001} and the fact that a star's luminosity is roughly proportional to the cube of its mass, it is easily shown that the stellar mass in galaxies is dominated by the low-mass stars, while the stellar luminosity is dominated by the high-mass stars. Wu05 and Wu10 suggested that the same low- and high-mass comparison for stars does not necessarily apply to dense gas mass and IR luminosity. Wu10 compared the HCN luminosity to the virial mass (\mvir), calculating the latter from the HCN spatial size and the linewidth of C$^{34}$S~(5--4). The power-law relation for \liriras~$\propto$~\mvir$^\gamma$ might be expected to have $\gamma$ near 2 since \liriras~$\propto~D^2$ and \mvir~$\propto~D$, where $D$ is the distance to the source. However, empirically Wu10 find $\gamma$ is $\sim$1. \mvir\ can be considered as a measure of the dense gas mass of a clump, \mdense. From these relations, they suggested that constant \lir/\lhcn\ is simply a reflection of a constant \funit.


The infrared luminosity of the Milky Way is estimated to be \lir\ $\approx 1.5 \times 10^{10}~L_\sun$ \citep{Cox1986}. While there is considerable difficulty in measuring the infrared luminosity of the Milky Way since it cannot be observed externally, this number is reasonable since NGC~891, an edge-on galaxy that strongly resembles the Milky Way, has a similar IRAS infrared luminosity of $2.6 \times 10^{10}~L_\sun$ (GS04). If the Milky Way follows the \gsr, \lhcn\ would be $\sim$10$^{7.3}$~\Kkmspcsq. For HCN, the mean and median MALT90 clump luminosity is \lhcn~=~10$^{2.1}$~\Kkmspcsq\ (Figure \ref{gaosolomon_IRASimfit_HCNonly}). If all high-mass clumps are approximately this luminosity, in order for high-mass clumps to sum up to the entire Milky Way \lir, $\sim$150,000 high-mass clumps are needed.




Alternatively we can estimate the number of high-mass clumps in the Galaxy if we assume that all dense gas in the Milky Way is contained within these clumps. \citet{Battisti2014} found that the ratio between the dense gas mass, \mdense, and the total molecular gas mass, $M_{\rm{molecular}}$, is $M_{\rm{dense}}/M_{\rm{molecular}} = 0.07^{+0.13}_{-0.05}$. The molecular gas mass of the Milky Way is $\sim$$2-3\times10^9~M_\odot$ \citep{Combes1991}, suggesting a dense gas mass of $\sim$$2\times10^8~M_\odot$. A typical high-mass star-forming clump mass is $\sim$1000~$M_\odot$ (e.g., Y. Contreras et al. in preparation), suggesting that if all dense gas are in these high-mass clumps, there would be 200,000 of these clumps. This estimate of the number of clumps is similar to the estimate based on the \gsr\ above. 

These two estimates suggest that if dense gas is primarily contained within high-mass star-forming clumps, there are on order $10^5$ high-mass star-forming clumps. However, the current estimates for the total number of such clumps in the Galaxy is much lower. For example, \citet{Zinnecker2007} suggest that only $\sim$5400 high-mass stars in the Milky Way are in their accretion phase, yet these stars presumably reside in the clumps that are the brightest with IRAS.  Moreover, the ATLASGAL survey is sensitive to all high-mass star-forming clumps with masses $>$200~$M_\sun$ out to a distance of 10 kpc (5$\sigma$ detections) \citep{Schuller2009,Jackson2013}, implying detections for almost all clumps $>$1000~$M_\odot$ in the surveyed region (420 sq. degrees over $-80^\circ < l < +60^\circ$). Only $\sim$10,000 \emph{total} clumps were detected by ATLASGAL \citep{Contreras2013,Csengeri2014,Urquhart2014} and many have masses below 1000~$M_\odot$ (Y. Contreras in preparation); while the survey did not completely sample the Galaxy, it surveyed the inner galaxy where most of the molecular mass is present. Therefore, the number of massive clumps $>$1000~$M_\odot$ is not likely to be significantly higher than $\sim$10,000. 

This analysis suggests that a significant amount of HCN emission from the Galaxy may not originate from dense, high-mass Galactic clumps. The same comparison can be drawn for the infrared luminosity. The mean and median MALT90 clump infrared luminosity is  \liriras~=~10$^{4.9}~L_\odot$ (Figure \ref{gaosolomon_IRASimfit_HCNonly}). For a Milky Way luminosity of \liriras\ = $1.5 \times\ 10^{10}$~$L_\odot$ \citep{Cox1986}, if dense clumps emit all the infrared luminosity at this average value, 190,000 clumps are needed, well above the expected number of Galactic high-mass clumps. Therefore, a significant amount of infrared emission from the Galaxy cannot originate from dense, high-mass Galactic clumps.

\begin{figure}[ht!]
\centering
\includegraphics[width=1\columnwidth]{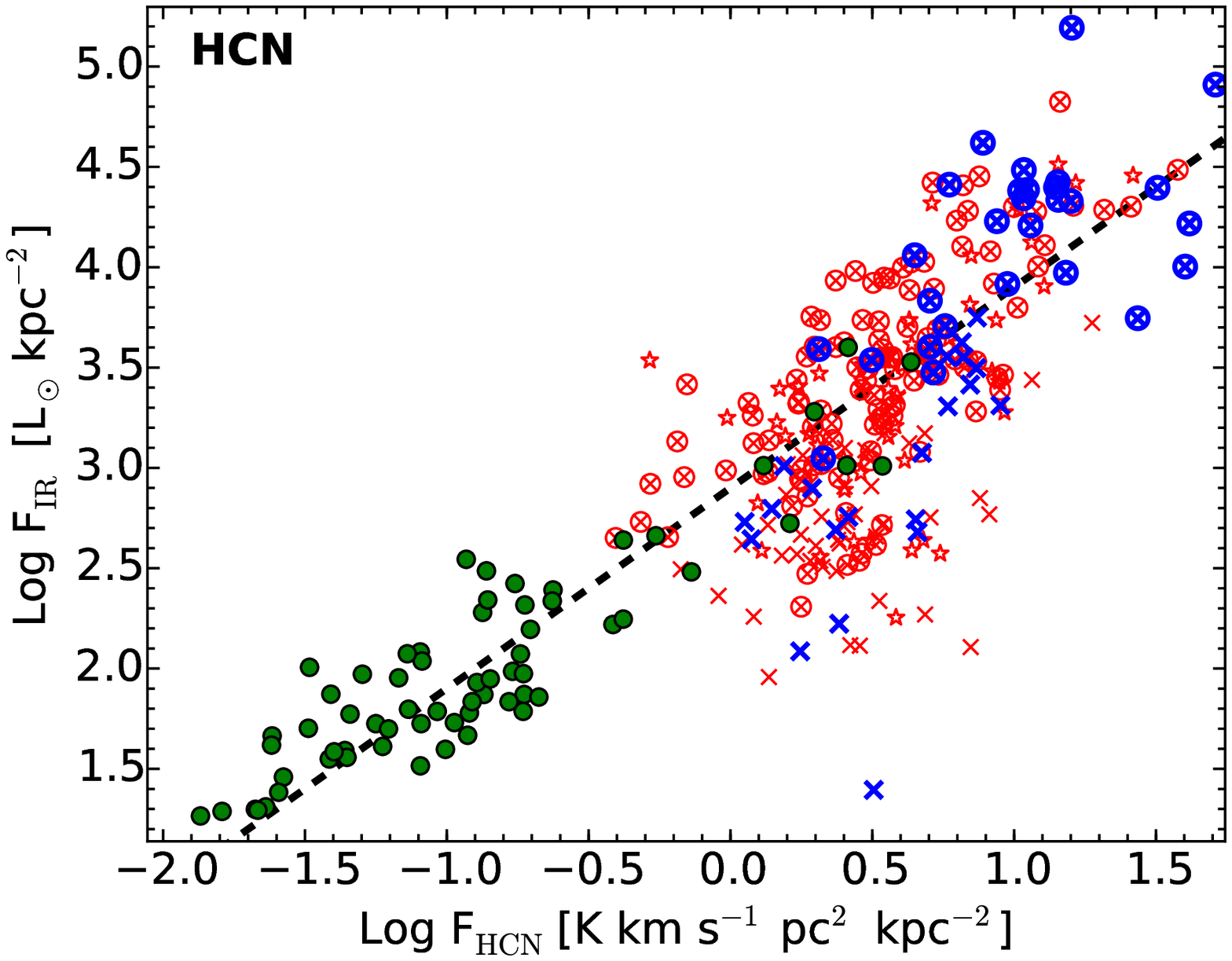}
\caption{Plot of \firiras\ and \fhcn\ for Galactic clumps and galaxies. Green circles are galaxies from GS04, blue crosses are Galactic clumps from Wu10  (with blue circles indicating if \lir~$>$~10$^{4.5}$~$L_\odot$), and red circles are Galactic clumps from MALT90 (with circles indicating if \lir~$>$~10$^{4.5}$~$L_\odot$ and stars indicating clumps with no kinematic distances). The dashed line shows the \gsr\ log\,$F_{\rm{IR}}$\,=\,1.00\,log\,$F_{\rm{HCN}}$\,+\,2.9. Units for \fhcn\ are in \Kkmspcsq$\,$kpc$^{-2}$ rather than \Kkms\ for easy conversion to luminosity. 
\label{gsrflux} }
\end{figure}

In Figure~\ref{gaosolomon_IRASimfit_HCNonly}, the scatter for the Galactic clumps is much larger than the scatter for galaxies from GS04. Part of the reason that the Galactic clumps may seem significantly correlated is due to the fact that both axes depend on the square of the distance which can cause the luminosities (\liriras\ and \lhcn) to be intrinsically correlated. Therefore, in Figure~\ref{gsrflux} we show the relation for fluxes \fhcn\ versus \firiras. Included in this plot are the GS04 galaxies, the Wu10 Galactic clumps, and the MALT90 Galactic clumps. We also have included in the plot an additional 51 MALT90 clumps that did not have valid distances but still have valid fluxes. For the GS04 galaxies, we multiplied the fluxes by the volumetric scale factor (1+$z$)$^{3}$. Note that the redshifts $z$ for most GS04 galaxies are very small, and the overall relation is unaffected by this factor. This plot shows that the GS04 sources follow a different slope than the Galactic clumps. Additionally, most of the Wu10 clumps with luminosities over 10$^{4.5}~L_\odot$ lie above the best fit line for the GS04 galaxies. This plot indicates that the extension to Galactic clumps may not be as tight of a relation as initially thought.



In summary, if the \gsr\ can be interpreted as beam-averaging typical high-mass star-forming clumps, then the number of clumps available in a galaxy appears to be insufficient by about an order of magnitude. Henceforth, we will call this mismatch between the estimated number of high-mass star-forming clumps and total luminosity of the Milky Way as the ``clump discrepancy." Moreover, the \firiras\ versus \fhcn\ plot (Figure~\ref{gsrflux}) provides evidence that the extension of the relation down to Galactic clumps may be poorer than suggested by the luminosity--luminosity plot. We investigate possible explanations for this clump discrepancy and the possible origin of the \gsr\ in the following subsections and summarize at the end.

\subsubsection{Investigating whether the Brightest Clumps or the Central Molecular Zone (CMZ) Dominate the IR and HCN Luminosities}\label{largecmz}

A possible reason for the apparent clump discrepancy is that the most luminous clumps dominate the luminosity of entire galaxies. For example, if the Galaxy contained 1000 clumps with an average \lhcn\ of $10^{4.3}$~\Kkmspcsq, the clump discrepancy would easily be accounted for. For the 160 clumps with valid HCN luminosities, the brightest clump has \lhcn~$\approx 10^{3.5}$~\Kkmspcsq. Approximately 6000 of these clumps can account for the total estimated \lhcn\ for the Milky Way ($\sim$10$^{7.3}$~\Kkmspcsq). Of the 160 IRAS-matched MALT90 clumps with valid HCN luminosities, only 5 clumps have  \lhcn~$>$~1000~\Kkmspcsq. Therefore, if the MALT90 sample accurately samples the high-mass clumps of the Galaxy (as it should, given the longitude coverage), the brightest clumps cannot dominate the Galactic HCN luminosity. Likewise, since the $median$ ratio \lir/\lhcn\ lies approximately on the \gsr\ for these clumps, they cannot dominate the Galactic IR luminosity either.


No kinematic distance for a MALT90 clump has been calculated for sources within $\pm$10$^\circ$ of the Galactic Center since a small change in velocity can change distances dramatically. However, toward the center of the Galaxy, the so-called CMZ, a large quantity of molecular gas exists. About one quarter (95 of 405) of MALT90 clumps that coincide with an IRAS source lie within $10^\circ$ of the Galactic Center. We therefore investigate whether the CMZ can contribute significantly to the total Galactic \lhcn, and if so, whether the emission comes directly from high-mass star-forming clumps.


With the Five College Radio Astronomical Observatory (FCRAO) 14-m telescope, \citet{Jackson1996} mapped the central $4\fdg3 \times 0\fdg5$ of the Galaxy \citep[630 $\times$ 73~pc assuming a distance of 8.34~kpc,][]{Reid2014}. Within this area, \citet{Jackson1996} calculated the HCN luminosity as \lhcn~=~$10^{6.5}$~\Kkmspcsq, which is approximately one-sixth of the luminosity expected for the projected Milky Way's \lhcn\ of $10^{7.3}$~\Kkmspcsq. Therefore, while the CMZ does not dominate the total Galactic \lhcn, the CMZ contributes a significant fraction (about one-sixth) to the total Galactic \lhcn. Within the same area there are 352 MALT90 clumps (J. Rathborne et al. in preparation) and 821 ATLASGAL clumps \citep{Csengeri2014}. If we assume there are indeed 821 high-mass star-forming clumps in this region, the average luminosity of a clump must be $10^{3.6}$~\Kkmspcsq\ in order to produce the total CMZ \lhcn. If we assign a Galactic Center distance of 8.34~kpc to all MALT90 clumps within 10$^\circ$ of the Galactic Center, the maximum \lhcn\ for any individual clump (excluding \_A and \_B clumps in Table~\ref{tab:main}) is $10^{3.26}$~\Kkmspcsq\ (AGAL351.161+00.697\_S). Since the highest HCN luminosity toward a Galactic Center clump is lower than the average HCN clump luminosity needed for clumps to dominate the total CMZ's luminosity, areas outside these clumps must provide a significant contribution to the HCN luminosity. Similarly, \citet{Cox1989} find that the total CMZ \emph{infrared} luminosity is dominated by extended areas. \citet{Cox1989} calculated a total infrared luminosity of $10^9~L_\odot$ in an area of $3^\circ \times 2^\circ$ about the Galactic Center and found that the compact sources only accounting for 10\% of this luminosity.

We quantify \lir\ and \lhcn\ within the same area of the CMZ. We use data from the Mopra CMZ 3\,mm Band Survey \citep{Jones2012}, which covers the CMZ in a box with dimensions of 2\fdg5~$\times$~0\fdg5 and centered at Galactic coordinates $l=0\fdg5$ and $b=0^\circ$. We use this solid angle for calculating the CMZ \lir\ and \lhcn. The available maps provide the antenna temperatures $T_A^*$ in the CMZ, i.e., the maps have not been corrected for the beam efficiency. For HCN, we first integrate over velocity to create an integrated intensity (moment 0) map. We then integrate over solid angle via \lhcn~$=D^2 I \Omega / \eta_{\rm{mb}}$, where $I$ is the velocity-integrated intensity and $\Omega$ is the solid angle. The main beam efficiency, $\eta_{\rm{mb}}$, was taken to be 0.5, and the distance, $D$, was taken to be 8.34~kpc. $I \Omega$ is equivalent to the sum of the pixels in the integrated intensity map times the solid angle subtended by a pixel. In the same box as this survey, we also calculated \lir\ as determined from IRAS using Equation~\ref{eqliriras}. For the area covered by the Mopra CMZ 3\,mm Band Survey, \lhcn~=~$10^{6.24}$~\Kkmspcsq\ and \lir~=~$10^{8.54}~L_\odot$. 

Figure 1 shows these derived CMZ IR and HCN luminosities. The data-point lies significantly below the regression fit from the \gsr, with a factor of 4.0 less than the expected \liriras\ given its \lhcn. To put this significance in perspective, only two of 65 GS04 galaxies is offset from the regression line by more than a factor of 3. The CMZ is very turbulent compared to the rest of the Galactic plane, which causes the Jeans mass to be much higher in the CMZ. Therefore, it is expected and known \citep[e.g.,][]{Longmore2013} that there will be a surplus of gas per unit of star formation, i.e., a smaller ratio of \lir/\lhcn. This discrepancy in \liriras\ in the CMZ would suggest that in order for the Milky Way to lie on the best fit line of the \gsr, on average the rest of the Milky Way must have a higher ratio of \lir/\lhcn\ than determined by the \gsr.  
For example, if the CMZ makes up one-sixth of the entire Milky Way's \lhcn\ and lies a factor of four below the \gsr, the rest of the Milky Way has to be 15\% above the \gsr\ in order for the entire Milky Way to follow the relation exactly. 
Different galaxies likely have different fractional quantities of HCN within their own CMZ, and these areas are subject to different conditions than the plane of a galaxy. Indeed, not all GS04 galaxies lie exactly on the regression fit, and it is conceivable that different quantities of gas and conditions within the central molecular zones of each galaxy can cause scatter in the \gsr. Future models probably should account for parameter differences toward the centers of galaxies (e.g., higher turbulence) when modeling the \gsr.

In summary, the brightest clumps cannot dominate the Galactic \lhcn\ and \liriras\ luminosities and thus account for the clump discrepancy. The CMZ, however, makes up a significant fraction ($\sim$10--20\%) of both the HCN and IR luminosity of the Milky Way, but the solid angle covered by the Mopra CMZ 3\,mm does not follow the \gsr. Moreover, extended emission outside dense clumps appears to add a significant (and perhaps a dominant) component to the CMZ HCN and IR emission.


\subsubsection{Investigating whether the Low-mass Star-forming Clumps Dominate the IR and HCN Luminosities}\label{lowmass}

The apparent clump discrepancy could be resolved if low-mass clumps rather than high-mass clumps dominate a galaxy's total \liriras\ and \lhcn. Indeed, the results in this paper primarily focus on bright, high-mass infrared clumps that are detected with IRAS. In this subsection, first we will address whether these low-mass clumps can dominate the total Galactic \lhcn, and then we will address whether they can dominate the total Galactic \lir. The Galaxy's total \lhcn\ and \lir\ from low-mass star-forming clumps depend on the clump mass function (ClMF) and the relation between the mass of the clump and each luminosity.

We assume that the ClMF takes the typical form d$N$/d$M~\propto~M^{-\delta}$. The empirical value for $\delta$ vary significantly \citep[values of 1.4--2.4, e.g.,][]{Elmegreen1996,Kramer1998}, although the more recent studies find $\delta \gtrsim 1.8$ \citep[e.g.,][]{Schneider2004,Reid2006,Pekruhl2013}. In the Carina nebula complex, \citet{Pekruhl2013} found that for clumps with masses between $\sim$50 and 3000~$M_\odot$, $\delta = 1.95 \pm 0.07$. $\delta$ depends on the input parameters and the clump finding algorithm, and \citet{Pekruhl2013} showed that using different temperatures or algorithms could result in values of $\delta$ from $1.89 \pm 0.06$ and $2.15 \pm 0.08$. Observations and simulations by \citet{Reid2006} and \citet{Reid2010}, respectively, found that the ClMF index is similar to the \citet{Salpeter1955} stellar IMF index, i.e., $\delta = 2.35$. We will consider both $\delta = 1.8$ and 2.35 for our analysis.

The relation between \lhcn\ and the clump mass is also not well determined, especially in the low-mass regime, but observations from Wu10 suggest that \lhcn\ may be directly proportional to the virial mass \mvir. The Wu10 sample primarily consists of high-mass clumps (\mvir~$\sim$~10$^2$ to $10^5$~$M_\odot$) but two low-mass clumps (\mvir~$\sim$~30 and 100~$M_\odot$) are also included and follow the same relation. We will assume \lhcn\ is proportional to $M$ for our calculations.

Based on the ClMF and the \lhcn--\mvir\ relation, we now determine the fractional contribution to a galaxy's total \lhcn\ from low-mass and high-mass star-forming clumps. Clumps above $\sim$200~$M_\odot$ are likely to form high-mass stars \citep[e.g.,][]{Jackson2013}. We assume the ClMF extends to the highest mass clumps with masses of $10^5~M_\odot$ \citep[e.g., G0.253+0.016, also known as the Brick,][]{Longmore2012}. The fractional \lhcn\ contribution from low-mass and high-mass star-forming clumps is strongly affected by our choice of the lower mass-limit of the ClMF. The simulation by \citet{Reid2010} suggests that $\delta$ is constant for clumps down to $\sim$2~$M_\odot$. Observations are not typically complete enough to sample this low-mass regime, but \citet{Pekruhl2013} showed that $\delta$ appears to at least be constant down to $\sim$50~$M_\odot$. We will assume that the ClMF extends down to 2~$M_\odot$. Assuming $\delta = 1.8$, clumps between 2 and 200~$M_\odot$ contribute approximately 25\% of the HCN luminosity as clumps between 200 and $10^5$~$M_\odot$. Alternatively assuming $\delta = 2.35$, clumps between 2 and 200~$M_\odot$ contribute approximately 4.5 times the HCN luminosity than clumps between 200 and $10^5$~$M_\odot$.

Given the uncertainties in this calculation, we cannot definitively determine whether high-mass or low-mass star-forming clumps contribute more significantly to a galaxy's total \lhcn. However, we have shown that it is certainly feasible that low-mass star-forming clumps ($<$200~$M_\odot$) can contribute up to a factor of $\sim$5 more to the total \lhcn\ of a galaxy than high-mass star-forming clumps ($>$200~$M_\odot$). Therefore, if high-mass star-forming clumps contribute approximately one tenth of a galaxy's total \lhcn\ (such as our Galaxy, beginning of Section \ref{gsrHCN}), low-mass star-forming clumps could contribute almost half of a galaxy's total \lhcn. If high- and low-mass star-forming clumps combined contribute $\sim$60\% to the Galaxy's \lhcn, and the Galactic Center's \lhcn\ luminosity is added (approximately one sixth of the Galaxy's total \lhcn, Section \ref{largecmz}), $\sim$75\% of the Galaxy's total HCN luminosity is accounted for.

Although low-mass star-forming clumps may dominate a galaxy's total \lhcn, they are not likely to dominate the total galaxy's \lir. According to Wu10, clumps with luminosities \lir\ are approximately directly proportional to \mvir\ for \lir~$>$~10$^{4.5}$~$L_\odot$ and \mvir~$>$~500~$M_\odot$. Clumps below $\sim$300~$M_\odot$ tend to have an order of magnitude less \lir\ per unit \mvir\ as those with \mvir~$>$~500~$M_\odot$. Therefore, low-mass star-forming clumps probably do not dominate a galaxy's total \lir. The fact that low-mass star-forming clumps may dominate \lhcn\ but not \lir\ indicates an excess of HCN per unit of infrared luminosity for low-mass star-forming regions. This excess of HCN emission is seen for clumps less than \lir~$<$~10$^{4.5}$ in Figures \ref{gaosolomon_IRASimfit_HCNonly} and \ref{gsrflux}. These clumps are a mix of low-mass star-forming clumps and clumps at a younger evolutionary stage; the latter will be addressed in Section \ref{sec:evol}.

To further characterize the possible contribution of low-mass star-forming clumps to the Galactic HCN luminosity, we consider the well-known low-mass star-forming region Serpens Main. The CARMA interferometer has publicly available interferometric HCN(1--0) data of this region at 7$\farcs$6 resolution \citep{Lee2014}. These observations used the CARMA23 mode which recovers the zero-spacing flux. Assuming a distance of 415~pc \citep[VLBA parallax observations,][]{Dzib2010}, Serpens Main consists of two sub-clumps of 97~$M_\odot$ and 144~$M_\odot$ \citep{Olmi2002}. From IRAS, we calculate \liriras\ luminosities of 60 and 123~$L_\odot$, respectively, resulting in a total \liriras\ of 183~$L_\odot$. Given \liriras\ of Serpens Main, the \gsr\ predicts \lhcn\ to be 0.23~\Kkmspcsq. However, fitting an elliptical Gaussian to the HCN emission of Serpens Main results in \lhcn~=~180~\Kkmspcsq\ for the entire complex -- almost 3 orders of magnitude higher than expected by the \gsr. Only $\sim$10$^5$ low-mass star-forming clumps like Serpens Main are required to produce the HCN luminosity of the entire Milky Way, which is certainly a plausible number of these clumps given that there may be approximately $10^4$ high-mass star-forming clumps (Section \ref{clumpsum}). 

In summary, while low-mass star-forming clumps are not likely to contribute a significant fraction to the total Galactic \lir, these clumps may contribute significantly to the total Galactic \lhcn.

\subsubsection{Investigating whether Subthermal Emission of HCN Dominates the HCN Luminosity}\label{subthermal}
The apparent clump discrepancy could be resolved if diffuse HCN emission not associated with dense clumps contributes significantly to the total HCN luminosity of the Milky Way. The HCN rotational level $J=1$ is easily populated for even cold gas since the temperature difference between the $J = 1$ and the $J =0 $ ground state is only 4.3~K. Although collisional excitation plays the most important role for HCN emission in regions above the HCN critical density, below this density HCN {radiative excitation dominates}. Excitation below the critical density is typically called subthermal excitation because the excitation temperature is lower than the kinetic temperature. All lines observed in the MALT90 survey have critical densities $>$10$^5$~cm$^{-3}$, and thus these molecular transitions are ideal for locating dense clumps associated with star formation. Outside of these dense clumps, HCN emission is expected to be primarily subthermal. The subthermally emitting HCN molecules are still subject to collisions, allowing for HCN(1--0) emission above the background temperature.

\citet{Helfer1997} observed CO(1--0) and the high-density tracers \hcnoz\ and CS(2--1) in an unbiased survey of the Galactic plane at $\sim$1$\arcmin$ resolution with the NRAO 12~m telescope. Between Galactic longitudes $l=15.5^\circ$ and $l=55.5^\circ$, they pointed the telescope in equally spaced increments of $1^\circ$. For almost every pointing, they detected all three lines. The integrated intensity ratios $I_{\rm{HCN}}$/$I_{\rm{CO}}$ and $I_{\rm{CS}}$/$I_{\rm{CO}}$ varied significantly from pointing to pointing. Each pointing generally detected lines with multiple velocity components, and the velocity components were typically at consistent velocities for all three molecular lines. \citet{Helfer1997} also mapped several GMCs and detected extended emission in CO(1--0), but only the dense clumps were detected in \hcnoz\ and CS(2--1). The typical integrated intensity ratios  in their unbiased survey is $I_{\rm{HCN}}$/$I_{\rm{CO}} = 0.026$ and for clumps detected in GMCs, the ratio is $I_{\rm{HCN}}$/$I_{\rm{CO}} = 0.1$. Since $I_{\rm{HCN}}$/$I_{\rm{CO}}$ is much larger toward dense clumps than the 1$^\circ$ separated locations along the Galactic plane, \citet{Helfer1997} proposed that the HCN emission is likely to be subthermal with a typical ratio $I_{\rm{HCN}}$/$I_{\rm{CO}} = 0.026 \pm 0.008$. For NGC~891, a galaxy considered very similar to the Milky Way, GS04 measured $I_{\rm{HCN}}/I_{\rm{CO}}$ to be 0.024, which is almost identical to the \citet{Helfer1997} measurements for the Galactic plane. Therefore, if the unbiased survey of the Galactic plane is primarily detecting subthermal HCN emission, it is likely that the HCN luminosity in GS04 galaxies can also be dominated by subthermal emission. 

Conversely, the model by \citet{KrumholzThompson2007} showed that for \hcnoz\ emission, GMCs do not emit a significant fraction of their HCN luminosities at densities $n \lesssim 10^4$~cm$^{-3}$; instead, most of the luminosity is emitted from gas near the critical density. This model suggests that purely subthermal excitation does not account for the dominant emission of a galaxy's HCN luminosity. Although \citet{KrumholzThompson2007} made some simplified assumptions (e.g., densities follow a log-normal probability distribution function in molecular clouds), there appears to be a clear discrepancy between the results of \citet{Helfer1997} and \citet{KrumholzThompson2007}.

\begin{figure*}[ht!]
\begin{center}
\includegraphics[width=1\textwidth]{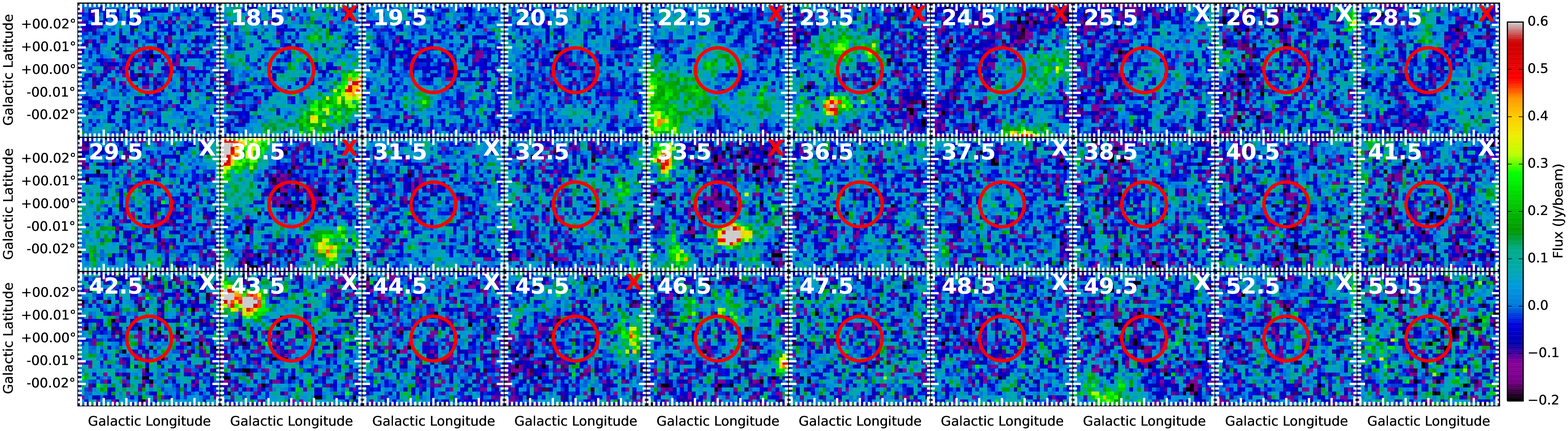}
\end{center}
\caption{ATLASGAL \citep{Schuller2009} 870~$\mu$m emission (color-scale, 19$\arcsec$ resolution) of the fields surveyed in \citet{Helfer1997}. The white number in the top left of each panel indicates the Galactic longitude of the pointing. The 71$\arcsec$ HCN beam is shown as a red circle in each panel. In general, each pointing is separated by a degree, although there are several pointings that \citet{Helfer1997} excluded due to poor data quality.  In the top right corner of each figure, no ``X" indicates an HCN peak radiative temperature $T^*_R < 0.05$~K, a white ``X" indicates $0.05$~K~$< T^*_R < 0.10$~K, and a red ``X" indicates $T^*_R > 0.10$~K. Note that the noise in each ATLASGAL map is not always the same, e.g., the noise for the $l=55.5^\circ$ map is higher than the other fields. Pointings that show significant HCN emission in \citet{Helfer1997} typically have significant nearby 870~$\mu$m emission.
}
\label{helferatlasgal} 
\end{figure*}

Given this apparent discrepancy between \citet{Helfer1997} and \citet{KrumholzThompson2007}, we further inspect the \citet{Helfer1997} study to analyze whether their data unequivocally show that the HCN emission from their Galactic plane survey is primarily subthermal. In Figure~\ref{helferatlasgal} we plot the 870~$\mu$m continuum data from ATLASGAL toward every \citet{Helfer1997} pointing. The red circle in each panel shows the FWHM of the NRAO 12~m beam (71$\arcsec$). In the top right of each figure, we indicate whether the HCN peak flux is  $T^*_R < 0.05$~K (no X), $0.05$~K~$< T^*_R < 0.10$~K (white X), or  $T^*_R > 0.10$~K (red X). While the integrated HCN intensity would give a better sense of the strongest HCN emission, this was not available in \citet{Helfer1997}. In general, pointings with higher HCN peak fluxes have significant 870~$\mu$m continuum flux, especially fields that have HCN $T^*_R > 0.10$~K (red X). Note that even though HCN emission may not be coincident within the FWHM of the beam, the beam is Gaussian, allowing for the strong emission bordering the FWHM of the beam to contribute significantly to the detected flux for the HCN pointing.

The \citet{Helfer1997} argument that this HCN emission is subthermal because the $I_{\rm{HCN}}$/$I_{\rm{CO}}$ ratio is significantly different to the ratio observed in clumps may be errant. The exact same result would be found if the beams are instead not centered on the clumps (i.e., the HCN peaks) since CO emission in clouds is more uniform than HCN emission. This indeed appears to be the case since Figure~\ref{helferatlasgal} shows that the \citet{Helfer1997} pointings are typically offset from the continuum peaks. By comparing Figure~5 of \citet{Helfer1997} with Figure~\ref{helferatlasgal} here, it is evident that the beams more centered on strong 870~$\mu$m continuum emission (e.g., $l=22\fdg5$) have much higher $I_{\rm{HCN}}$/$I_{\rm{CO}}$ values than those with 870~$\mu$m emission along the border of the beam (e.g., $l=30\fdg5$). Moreover, the 870~$\mu$m continuum maps of the ATLASGAL survey appear filamentary, which likely indicates locations of elongated, dense structures rather than areas of subthermal emission. The multiple HCN velocity components detected in \citet{Helfer1997} could simply be explained by multiple clumps at different distances; for example, the fields centered at longitudes 22\fdg5, 23\fdg5, 30\fdg5, and 33\fdg5 obviously have multiple clumps in the field that could be at very different distances. Moreover, for the spectra in \citet{Helfer1997}, the ratio $I_{\rm{HCN}}$/$I_{\rm{CO}}$ for velocity components within a single pointing vary significantly; for two clumps with different velocities, this would be expected if the beam is centered more on one clump than the other. 

The integrated intensity ratio of the \citet{Helfer1997} Galactic plane survey ($I_{\rm{HCN}}$/$I_{\rm{CO}} = 0.026$) is still an important quantity even though there certainly is some contamination by clumps; since the survey is unbiased, the intensity ratio represents the expected intensity when beam-averaging the clumps throughout the plane of a galaxy similar to the Milky Way. This could explain why a similar integrated intensity ratio, i.e., $I_{\rm{HCN}}/I_{\rm{CO}} = 0.024$, was found for NGC~891 in GS04.

Although the \citet{Helfer1997} Galactic plane survey was sometimes coincident with clumps along the line of sight, a significant contribution to a galaxy's total HCN emission may still come from subthermal HCN emission. For example, \citet{McQuinn2002} mapped a two square degree area in the Galactic plane with the high-density tracer CS(2--1) and found $3\sigma$ detections for $\sim$75\% of the solid angle covered. 85\% of the 3$\sigma$ detections comes from areas with no obvious clumps, suggesting these regions have subthermal-emitting gas. This supposed subthermal emission contributes $\sim$65\% to the total intensity within the two square degree map and thus dominates the emission in the region. Moreover, many extragalactic observations have suggested that subthermal emission of a high-density tracer can contribute significantly to the line luminosity of an entire galaxy \citep[e.g.,][]{Papadopoulos2007,Aravena2014}.


In summary, measurements of previously reported Galactic subthermal emission from \citet{Helfer1997} is probably contaminated by clumps. The ratio of Galactic plane intensities found by \citet{Helfer1997}, $I_{\rm{HCN}}$/$I_{\rm{CO}} = 0.026$, probably does not quantify the Galactic HCN contribution from only subthermal emission but rather the expected ratio observed in a plane of a galaxy similar to a Milky Way. However, based on other studies, it is possible that subthermal emission can contribute significantly to a galaxy's total \lhcn. Diffuse gas with densities below the HCN critical density (i.e., subthermally excited) may dominate the total HCN luminosities from galaxies due to the much larger solid angle occupied by the diffuse gas compared to the solid angle occupied by dense clumps.


\subsubsection{Investigating whether the Extended Emission due to the Interstellar Radiation Field (ISRF) Dominates the Infrared Luminosity}
High-mass OB stars dominate the energy of the ISRF and the flux in IRAS bands. Cold dust ($\lesssim25~K$) is heated by the ISRF, warm dust ($\sim$30--40~K) is heated by young stellar populations, and hot dust ($\gtrsim$250~K) is heated by the general ISRF and OH/IR stars \citep{Cox1986}. Emission form all three dust components contribute significantly to the total IR flux. Given that the general ISRF includes contributions from both younger and older high-mass stellar populations, there may be significant contribution to the total Milky Way's IR luminosity from dust not heated by embedded high-mass young stellar objects (YSOs). Extended, more diffuse regions, may be heated by either the general ISRF or by photons escaping from young regions and thus may contribute a significant fraction to the total IR luminosity. Already we suggested this could be the case in the CMZ (Section~\ref{largecmz}).

\begin{figure*}[ht!]
\centering
\includegraphics[width=1\textwidth]{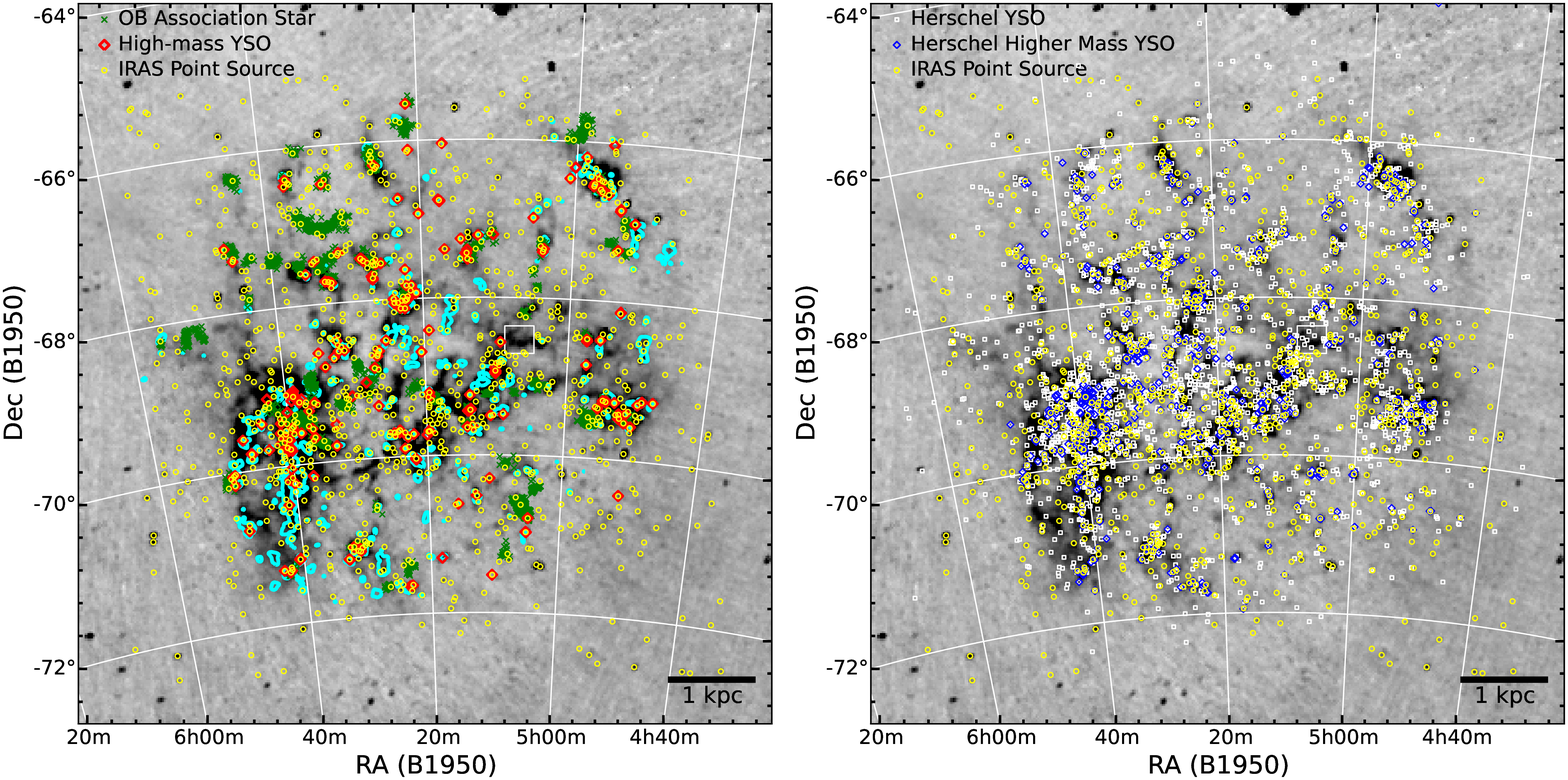}
\caption{IRAS 12~$\mu$m images (grayscale) of the Large Magellanic Cloud. Left: the locations of high-mass young stellar objects \citep{GC09} are shown with red diamonds and OB stars of known OB associations \citep{Lucke1970} are shown with green crosses. Sources from the IRAS Point Source Catalog are shown with yellow circles. Cyan contours are 3$\sigma$ CO(1--0) detections from NANTEN \citep{Fukui2008}, indicating locations where the column density is $N$(H$_2$)~$\gtrsim$~$8\times 10^{20}$~cm$^{-2}$. The white rectangle shows a location where there is an IRAS detection but no IRAS point source, no \citet{GC09} high-mass YSO, and no OB association stars. Right: the locations of known $Herschel$ YSOs are in white and higher-mass YSOs (i.e., sources with $Herschel$ luminosities $>$1000~\lsun) are in blue. Like the left panel, we also show the IRAS point sources and a white rectangle for an area of interest.
\label{LMC_IRAS} }
\end{figure*}

To understand the infrared contribution from high-mass YSOs and extended IR emission to a galaxy's total \lir, it is helpful to inspect an external galaxy. The Large Magellanic Cloud (LMC) is perhaps the best example of a nearby galaxy that can be studied at high enough spatial resolution to resolve individual stars and YSOs. The LMC is located at 50~kpc \citep{Feast1999} and is nearly face-on \citep[35$^\circ$ inclination with respect to the plane of sky,][]{vanderMarel2001}. We present the 12~$\mu$m IRAS image of the LMC in Figure~\ref{LMC_IRAS}. Overlaid on this image are the locations of $Spitzer$-identified high-mass YSOs \citep[the ``definite" and ``probable" YSOs identified in][]{GC09}, known OB associations \citep{Lucke1970}, and IRAS sources from the IRAS Point Source Catalog. Almost all \citet{GC09} high-mass YSOs are at locations with strong 12~$\mu$m IRAS emission, but many OB associations (i.e., high-mass star conglomerations that are not embedded) are not closely associated with any 12~$\mu$m emission. Conversely, there are many areas with strong 12~$\mu$m emission that do not have a high-mass YSO as identified by \citet{GC09}.

In the left panel of Figure~\ref{LMC_IRAS}, we also show the known GMCs in the LMC as probed by CO(1--0) with the NANTEN telescope \citep{Fukui2008}. This survey had a 3$\sigma$ detection limit for molecular clouds with molecular hydrogen column density of $N$(H$_2$)~$>$~$8 \times 10^{20}$~cm$^{-2}$ (assuming a conversion factor $X_{\rm{CO}}$ of $7\times10^{20}$ cm$^{-2}$ (K km s$^{-1}$)$^{-1}$). If we assume a minimum thickness of 1~pc for the GMCs along the line of sight (this value is conservatively low since the plane-of-sky extent of CO emission is typically much larger than this; Figure~\ref{LMC_IRAS}), a non-detection must have an average density within the NANTEN 2$\farcm$6 beam of $n_{\rm{H_2}} < 260~\rm{cm}^{-3}$. Nevertheless, there are large areas in the LMC that are bright with IRAS but have no detection of CO from NANTEN, indicating $N$(H$_2$)~$<$~$8 \times 10^{20}$~cm$^{-2}$ in these areas. This extended IR emission can reach scales of over 100~pc, and given the average column (and inferred volume) density at these scales, HCN will not be thermally excited.

In order to test whether this extended IR emission outside of the known CO(1--0) gas behaves according to the \gsr, we select an area with IRAS emission void of any \citet{Fukui2008} CO(1--0) cloud, IRAS source, \citet{GC09} YSO, and \citet{Lucke1970} OB association stars, as shown with the white box (size $20\arcmin \times 21\arcmin$, or $290 \times 305$~pc)  in both panels of Figure~\ref{LMC_IRAS}. The infrared luminosity in this box (using Equation~\ref{eqliriras}) is \liriras\ = $10^{6.6}~L_\odot$. If this region followed the \gsr, the HCN integrated intensity would be 2.1~\Kkms. The CO(1--0) survey with NANTEN has a $3\sigma$ detection limit of $\sim$1.2~\Kkms, yet CO, which is typically a factor of $\sim$10 to 100 times brighter than HCN \citep[e.g.,][]{Helfer1997}, was completely undetected in this region. Therefore, this extended IR emission does not behave according to the \gsr.

We compare \liriras\ of the entire LMC ($L_{\rm{IR,LMC,All}}$) and the sum of \liriras\ for all sources in the IRAS PSC ($L_{\rm{IR,LMC,PSC}}$), and we find that $L_{\rm{IR,LMC,PSC}}$ does not dominate $L_{\rm{IR,LMC,All}}$.  To calculate $L_{\rm{IR,LMC,All}}$, we sum the flux for each IRAS band of the LMC about a $7\fdg5 \times 7\fdg5$ box centered at R.A. (B1950) = $5^{\rm{h}}20^{\rm{m}}$ and decl. = $-68^\circ50\arcmin$. From Equation~\ref{eqliriras} we find that $L_{\rm{IR,LMC,All}}$ = 10$^{9.1}~L_\odot$. The sum of \liriras\ of all the IRAS point sources (shown as yellow circles in Figure~\ref{LMC_IRAS}) in the same box is $L_{\rm{IR,LMC,PSC}} = 10^{8.2}~L_\odot$. The IRAS point source catalog detects any LMC IR source with \liriras~$\gtrsim 10^{3.1}~L_\odot$, suggesting that all the high-mass star-forming regions (\liriras~$\gtrsim 10^{4.5}~L_\odot$) are detected. The sum of the point sources is only 12\% of the entire LMC IR luminosity. As discussed at the beginning of Section~\ref{gsrHCN}, this 12\% factor is similar to the fraction of \liriras\ that $\sim$10,000 high-mass clumps are expected to contribute to the Milky Way's total \liriras.

Since there are no known high-mass YSOs in some LMC extended IR regions, the IR emission probably comes from dust grains heated by the LMC's ISRF \citep{Cox1986}. The HERITAGE $Herschel$ survey (100--500 $\mu$m) of the LMC \citep{Meixner2013}  provides a much more complete catalog of LMC IR sources \citep{Seale2014} due to its higher sensitivity (especially for cold sources) and much higher resolution. In the right panel of Figure~\ref{LMC_IRAS}, we show the most probable YSOs in the LMC. Specifically, we chose all sources in the \citet{Seale2014} catalog that were considered as ``probable" or ``possible" YSOs as well as dust clump sources with luminosities $>$1000~$L_\odot$. Dust clumps that had luminosities $>$1000~$L_\odot$ have a very high chance of containing an embedded YSO since the luminosity of the dust clump cannot be explained by the LMC ISRF only \citep{Seale2014}. We also mark in Figure~\ref{LMC_IRAS} the ``higher mass" YSOs, which includes all \citet{Seale2014} sources (``probable" and ``possible" YSOs and dust clumps) with luminosities $>$1000~$M_\odot$. Note that a small fraction of the bright YSOs and dust clumps in \citet{Seale2014} had $Herschel$ fluxes that failed the SED fitting process and thus could not be labeled in the Figure as a ``higher mass" YSO. Most extended IR regions consist of at least one low-mass YSO, and these YSOs have insufficient luminosity to contribute significantly to the infrared luminosity of the entire cloud. This dust is more likely heated by the ISRF than embedded stars. 

This analysis casts doubt on the interpretation put forth by Wu05 and Wu10. They proposed that above \liriras~$>~10^{4.5}~L_\odot$, star-forming molecular clumps have a roughly constant value of \lir/\lhcn. The sum of the luminosities of large numbers of such clumps then produce the total \liriras\ and \lhcn\ of a galaxy, conserving the linear relation between \liriras\ and \lhcn. We have shown that a significant fraction of IR emission in the LMC comes from diffuse areas rather than IRAS clumps, and these areas probably do not follow the \gsr. A significant fraction of $L_{\rm{IR,LMC,All}}$ likely comes from ISRF. Similar studies have also calculated the IR contribution from different components in the Milky Way. \citet{Mezger1982} showed that $\sim$80\% of the Galaxy's far-infrared luminosity comes from low density regions, while only $\sim$10\%--20\% comes from more compact sources. The extended emission is heated by approximately equal fractions from the ISRF and diffuse emission from low-density areas of \ion{H}{2} regions \citep{Cox1989}. Based on previous studies and our analysis here, we suggest that extended emission does not have a characteristic value \lir/\lhcn\ and likely dominates the total infrared luminosity of galaxies.






\subsection{Summary of Origin of the Clump Discrepancy}
We introduce the idea of a clump discrepancy, i.e., if high-mass clumps in a galaxy are units that sum to produce the observed \lir/\lhcn\ of an entire galaxy, then the Milky Way needs a factor of $\sim$10 more high-mass star-forming clumps than are currently expected to exist to satisfy the \gsr. This suggests that the current estimates of the number of high-mass star-forming clumps in the Milky Way are either drastically wrong or some of the IR and HCN emission must originate outside high-mass clumps. We analyze four other possible origins of the missing \lir\ and \lhcn\ within the Milky Way:

\begin{itemize}
\item We first investigated whether the largest clumps or the CMZ could dominate \lir\ or \lhcn\ in the Milky Way. The former appears unlikely given the current known population of clumps in the Galaxy. The CMZ does not dominate these luminosities for the Milky Way either, but does contribute 10--20\% to these total luminosities \citep{Cox1989,Jackson1996}. Extended emission contributes significantly to the CMZ's IR and HCN luminosity.

\item We then investigate whether low-mass clumps can dominate  \lir\ or \lhcn\ in the Milky Way. While low-mass clumps do not contribute a significant fraction to \lir, it is feasible that low-mass clumps can dominate \lhcn. 

\item We also investigated whether subthermal HCN emission could dominate \lhcn. If the HCN detections from the \citet{Helfer1997} Galactic plane survey indeed represent subthermal emission of HCN, then subthermal HCN emission could dominate the total \lhcn. Our analysis of the \citet{Helfer1997} suggests that at least some of this HCN emission is not subthermal. Nevertheless, other observations suggest that there could be significant subthermal emission in both the Milky Way and other galaxies, and such emission could contribute significantly to a galaxy's entire HCN emission. 

\item Finally, we investigated whether extended IR emission can dominate \lir. Based on analysis of the CMZ, the LMC, and previous studies, it appears that extended IR emission can dominate \lir\ for a galaxy. Moreover, this extended emission does not appear to follow the \gsr. 

\end{itemize}

Based on these results, it appears that extended IR emission probably dominates the total \lir\ in galaxies. We are uncertain what dominates the HCN emission in galaxies, but we suggest low-mass star-forming clumps and/or subthermal emission could possibly dominate this emission. We discuss the implications of these results in Section~\ref{discussion}.


\section{Extending the Comparison Between the IR and Dense Gas Luminosity to Other Molecules}\label{gsrall}
\begin{figure*}[ht!]
\includegraphics[width=1\textwidth]{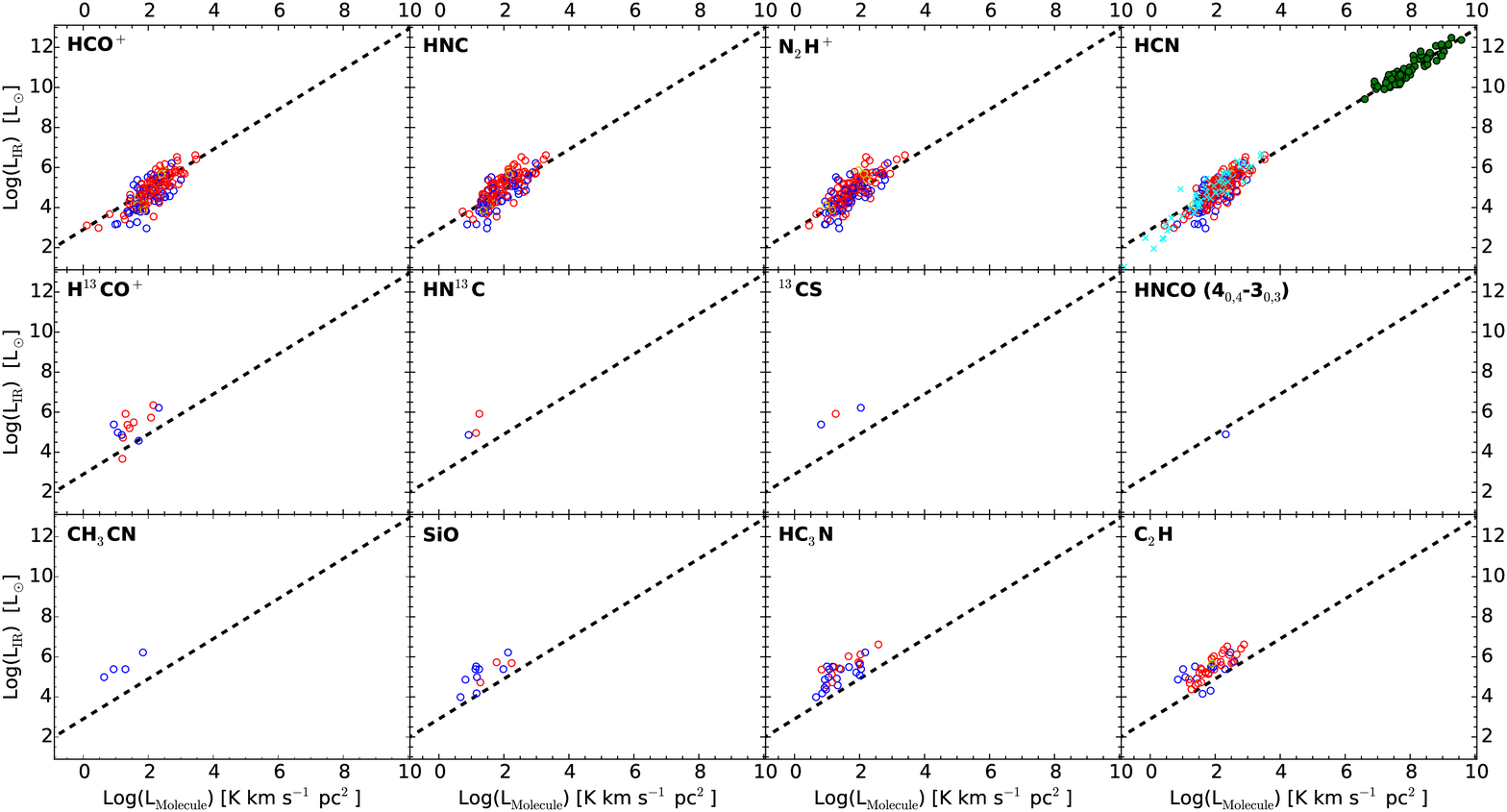}
\caption{\liriras\ versus \lmol\ for twelve different MALT90 molecular lines. As identified using $Spitzer$ near-IR images (Section~\ref{sec:evol}), red circles indicate clumps classified as \ion{H}{2} and Compact \ion{H}{2} regions, blue circles indicate clumps classified as Protostellar, and yellow circles are clumps with other classifications. Green circles are galaxies from GS04 while cyan crosses are from Wu10. The dashed line shows the \gsr\ log\,$L_{\rm{IR}}$\,=\,1.00\,log\,$L_{\rm{molecule}}$\,+\,2.9.
\label{gaosolomon_IRASimfit} }
\end{figure*}

\begin{deluxetable}{lccc}
\tablecolumns{4}
\tabletypesize{\footnotesize}
\tablewidth{0pt}
\tablecaption{IRAS Luminosity versus Line Luminosity, all \liriras\ luminosities \label{tab:fits}}
\tablehead{\colhead{Molecular Line} & \colhead{Slope} & \colhead{Intercept} & \colhead{$R^2$}
}
\startdata
HCO$^+$(1--0) & 1.16~$\pm$~0.08 & 2.38~$\pm$~0.17 & 0.60 \\
HNC(1--0) & 1.16~$\pm$~0.08 & 2.64~$\pm$~0.16 & 0.61 \\
N$_2$H$^+$(1--0) & 1.13~$\pm$~0.08 & 2.81~$\pm$~0.14 & 0.61 \\
HCN(1--0) & 1.21~$\pm$~0.07 & 2.32~$\pm$~0.15 & 0.66 \\
H$^{13}$CO$^+$(1--0) & 0.99~$\pm$~0.40 & 3.78~$\pm$~0.62 & 0.36 \\
SiO(2--1) & 0.95~$\pm$~0.27 & 3.85~$\pm$~0.39 & 0.56 \\
HC$_3$N(10--9) & 0.93~$\pm$~0.15 & 3.92~$\pm$~0.23 & 0.57 \\
C$_2$H(1--0)\tablenotemark{a} & 0.92~$\pm$~0.12 & 3.72~$\pm$~0.23 & 0.57 \\
\enddata
\tablecomments{Format of regression fit is $\mbox{log}(L_{\rm{IR}})$ = Slope $\times~\mbox{log}(L_{\rm{molecule}})$ + intercept. $R^2$ is the Pearson coefficient of determination.}
\tablenotetext{a}{See Table~\ref{tab:m90} for specific transition.} 
\end{deluxetable}

In  Figure~\ref{gaosolomon_IRASimfit} we present the relation between \lir\ and \lmol\ for the 12 molecular lines with significant numbers of detections by MALT90 (see Section \ref{sec:calcs}). The four molecular lines most often detected, \hcop, HNC, \nthp, and HCN (1--0), lie in the vicinity of the best fit regression line for the \gsr. Therefore, galaxies may also follow a Gao--Solomon type relation using the tracers \hcop, HNC, and \nthp. In general, datapoints for the other molecular lines (i.e., excluding \hcop, HNC, \nthp, and HCN) have higher \lir/\lmol\ ratios than that found by GS04. The datapoints for these molecular lines are expected to have higher ratios since they have similar critical densities but are significantly less abundant than HCN.\footnote{For \cth\ $N=1-0$, we only integrate the transition listed in Table \ref{tab:m90}; adding all $N=1-0$ transitions would increase these luminosities considerably.} In Table~\ref{tab:fits} we show the ordinary least squares (OLS) fits for log~\liriras\ versus log~\lmol\ for all transitions except HN$^{13}$C, $^{13}$CS, \hncofzf, and CH$_3$CN (these transitions have four or less datapoints). The OLS regression was chosen over other regression fit algorithms because the uncertainties for each datapoint are not well determined. Moreover, using an OLS regression allows for a direct comparison to GS04 OLS fits. We test the behavior of the OLS regression for each line by using the random resampling with replacement bootstrapping technique \citep[e.g.,][]{Simpson1986}. The bootstrap tests confirmed the uncertainties in the OLS fitted slope and intercept values and demonstrated that the OLS fit results are insensitive to uncertainties in source distances at the $\sim$30\% level. Fits from Wu05 and Wu10 exclude clumps with \liriras~$<10^{4.5}~L_\odot$, but we do not provide these fits because there is a large spread in the data. Specifically, with the luminosity cut, the spread of the data for the y-axis (\lir) becomes very similar to the spread of the x-axis (\lmol) and thus regression dilution bias significantly reduces the OLS slope.


\begin{figure*}[ht!]
\includegraphics[width=1\textwidth]{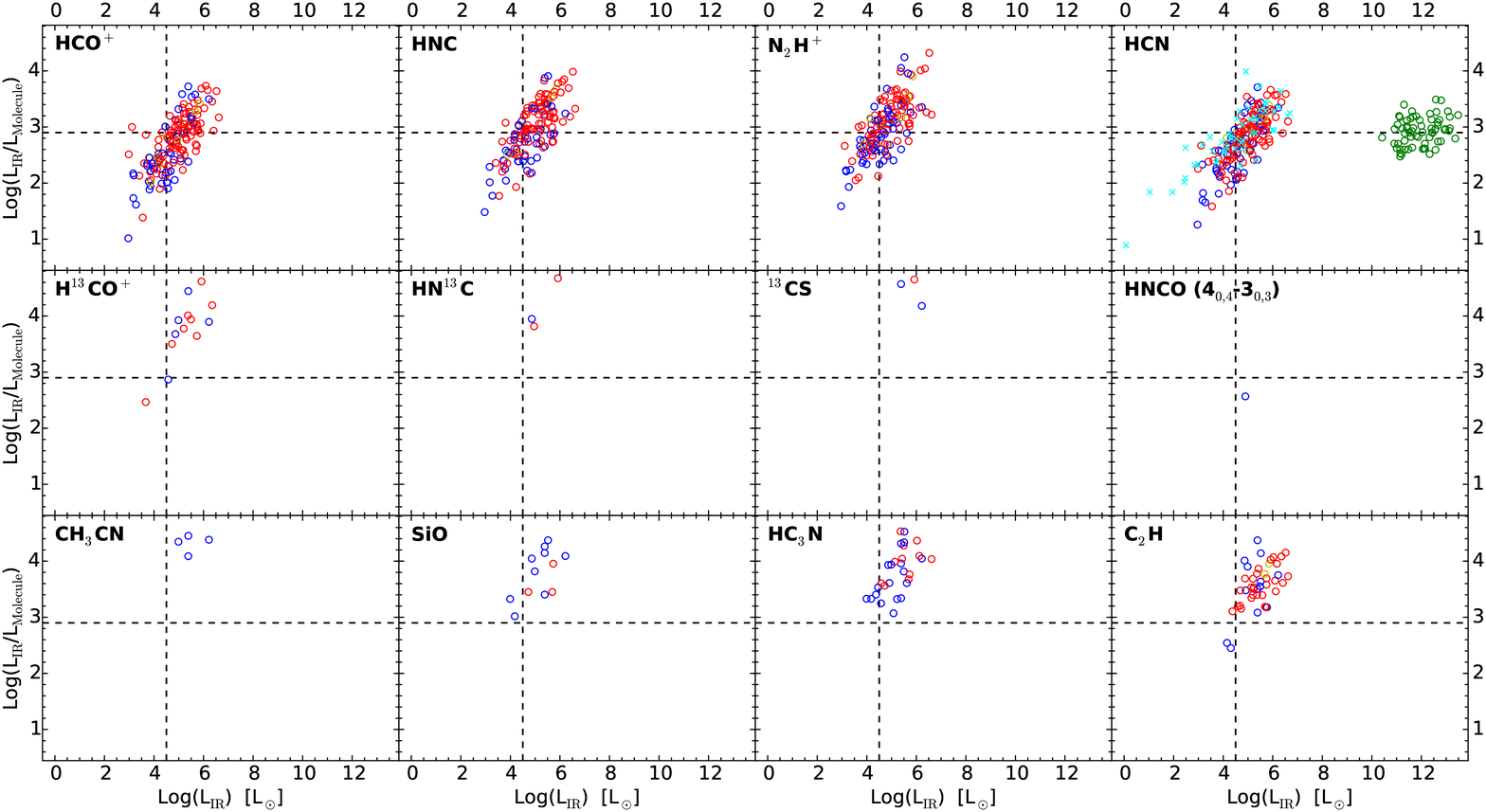}
\caption{Distance-independent \liriras /\lmol\ versus \liriras\ for different molecular lines. \liriras /\lmol\ are in units of \lsun$\,$(\Kkmspcsq)$^{-1}$.  Colors and markings are the same as Figure~\ref{gaosolomon_IRASimfit}. Dashed reference lines are shown for \liriras~$=10^{4.5}~L_\odot$ (vertical line) and the average ratio for GS04 galaxies \liriras /\lmol\ = 10$^{2.9}$ \lsun$\,$(\Kkmspcsq)$^{-1}$ (horizontal line). \liriras /\lmol\ decreases with decreasing \lir. There appears no evidence of a definitive drop in \lhcn\ at a particular value of \lir\ as proposed by Wu05 and Wu10.
\label{normalized} }
\end{figure*}

In Figure~\ref{normalized} we show the ratio \liriras/\lmol\ versus \liriras. \liriras/\lmol\ is independent of distance; therefore, this figure is similar to the \firiras\ versus \fhcn\ plot shown in Figure~\ref{gsrflux}. Wu05 and Wu10 used a similar plot with their data to argue that \liriras/\lhcn\ is constant for infrared luminosities \lir~$> 10^{4.5}~L_\odot$. They argue that below $10^{4.5}~L_\odot$ (approximately the luminosity of an \ion{H}{2} region), there is a sudden drop in \liriras/\lhcn. We do not see a sudden drop off at this luminosity; instead, we see the general trend that more luminous (and likely higher mass) infrared clumps have excess \liriras\ per unit \lhcn\ as compared to the less luminous clumps. This is consistent with our interpretation in Section~\ref{gsrHCN} suggesting that low-mass clumps have a significant excess \lhcn\ per unit \liriras. The trend may be due to the following: (1) More massive clumps are more likely to sample the high-mass regime of the initial mass function. Since the luminosity of stars scales with the third power of its mass, the dust of embedded clumps that host the most massive stars will absorb more luminosity which in turn will reemit in the infrared; (2) More massive clumps may have higher densities and thus the HCN may be optically thick, causing the luminosity to be suppressed; and/or (3) More infrared-luminous clumps are often \ion{H}{2} regions, and these clumps typically have lower column densities than protostellar clumps \citep{Guzman2015}. If these clumps are optically thin, a deficit in HCN(1--0) luminosity would exist for a given infrared luminosity.


For the four brightest lines, \hcop, \hnc, \nthp, and \hcn, the slope of the linear regression fit of log(\liriras/\lmol) versus log(\liriras) is $\sim$0.5, indicating some correlation between \liriras\ and \lmol. In other words, Figure~\ref{normalized} is not a plot of \lir\ versus \lir, which has an expected linear regression slope of 1.

\begin{figure*}[ht!]
\centering
\includegraphics[width=1\textwidth]{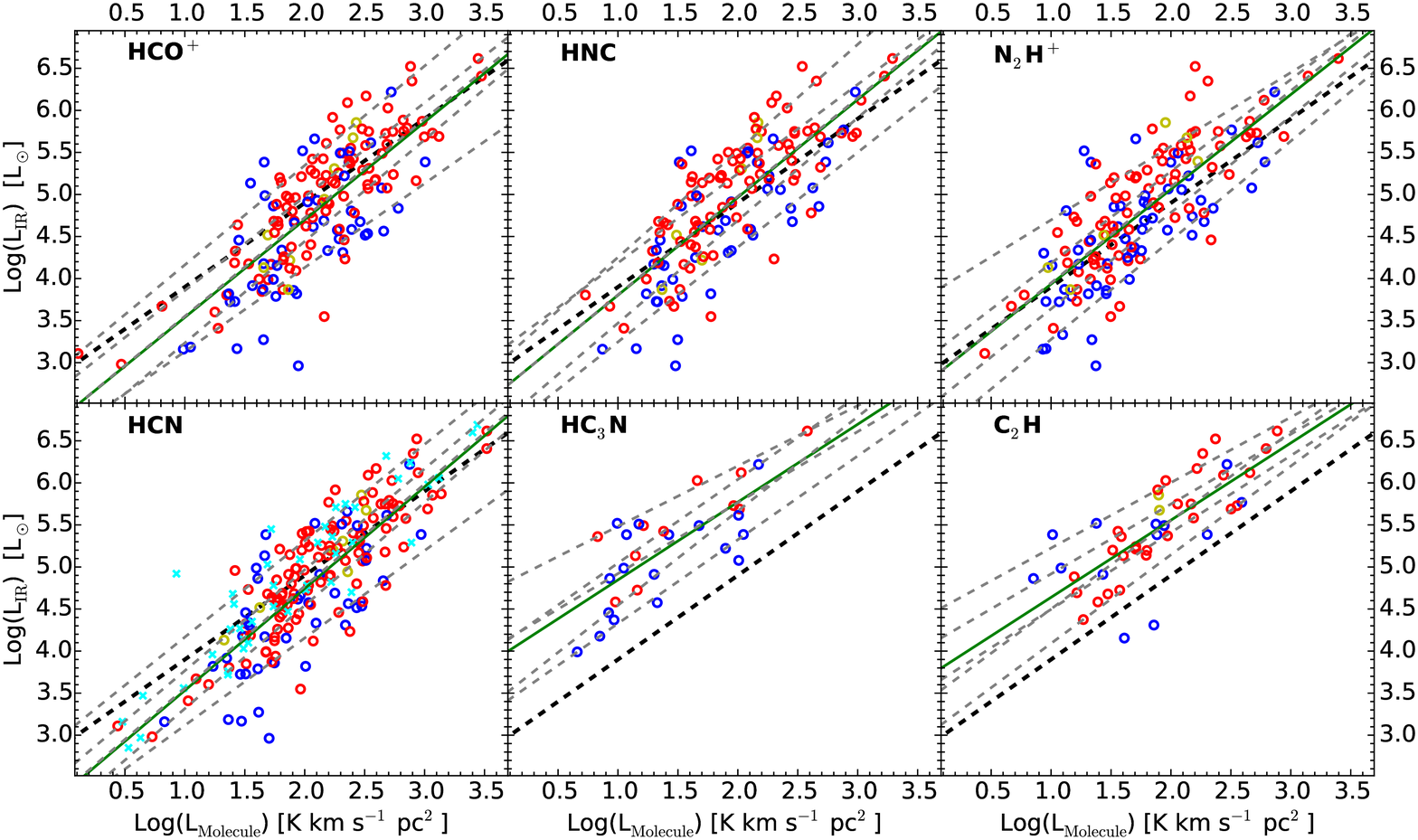}
\caption{The \lir\ and \lmol\ relation for the six molecular lines with the most MALT90 detections. Clumps classified as \ion{H}{2} Region clumps (compact and not compact) and Protostellar clumps are shown in red and blue, respectively, while clumps with other classifications are shown in yellow. The black dashed line shows the fit to the GS04 sources. The green line shows the least squares fit for each molecular line for the MALT90 clumps. Five gray dashed lines show the quantile regression fits for quantiles (bottom to top line) $q$ = 0.1, 0.3, 0.5, 0.7, and 0.9, respectively. Clumps from Wu10 for HCN are shown as cyan crosses and were not used in any of the fits.
\label{quantile} }
\end{figure*}

In order to analyze the clump-scale \gsr\ in more detail, we plot the six lines with the most detections,  \hcop, HNC, \nthp, HCN, HC$_3$N, and C$_2$H, in Figure~\ref{quantile}. In this Figure, we show the best fit line for the \gsr\ and the linear regression fit for the MALT90 clumps. \hcop\ and HCN lie slightly below the best fit line for the \gsr\ while \nthp\ and HNC lie on top of the line. A Kolmogorov-Smirnov test indicates that for MALT90 clumps, \lir/\lhcn\ and \lir/\lhcop\ could be drawn from the same distribution ($P$-value of 0.70), but  \lir/\lhnc\ and \lir/\lnthp\ likely come from a different distribution than \lir/\lhcn\ ($P$-values of less than 10$^{-6}$). The abundances and excitation parameters for these two molecular lines are not expected to be the exactly the same, and abundances of these lines can be affected by physical conditions of the region.



We also investigate the spread of the correlations between \liriras\ and the line luminosities for different molecules (\lmol) in order to discern which molecule shows the tightest correlation. In Table~\ref{tab:fits} we show the tabulation of the coefficient of determination, $R^2$, for each relation. Except for H$^{13}$CO$^+$, all lines with regression fits have similar $R^2$ values ($\sim$0.6) indicating that they all have a similar spread. In order to use a measurement that is less affected by outliers, we also plot the quantile regression fits \citep[e.g.,][]{Koenker2001}, for quantiles $q~=~0.1$, 0.3, 0.5, 0.7, and 0.9 in Figure~\ref{quantile} (where $q$ is the expected fraction of data below the $q$ quantile regression fit line). The dispersion of the quantile fits, particularly for the four brightest lines \hcop, HNC, \nthp, and HCN, are very similar. These four brightest lines have a \lir/\lmol\ difference of $\sim$2 orders of magnitude between quantiles $q=0.1$ and $q=0.9$.


\subsection{Clump Evolutionary Stage and the \lir\ and \lmol\ Relation}\label{sec:evol}
In the MALT90 catalog (J. Rathborne et al. in preparation), the evolutionary stage of each MALT90 clump has been classified based on visual inspection of mid-IR images from $Spitzer$ IRAC and MIPS \citep[3.6--24 $\mu$m,][]{Benjamin2003,Carey2009}. These clumps have been classified as Quiescent, Protostellar, Compact \ion{H}{2} region, \ion{H}{2} region, photodissociation region (PDR), or Uncertain. The classifications are indicated in Table~\ref{tab:main}. Details of the classification scheme are discussed in multiple MALT90 papers \citep[e.g.,][]{Hoq2013,Jackson2013,Stephens2015a}. Since IRAS only detects the brightest clumps, the clumps analyzed in this paper are primarily those classified as Protostellar, Compact \ion{H}{2} region, or \ion{H}{2} region. Protostellar clumps have compact 24 $\mu$m emission and enhanced 4.5~$\mu$m emission \citep[``green fuzzies,"][]{Chambers2009}. \ion{H}{2} regions typically appear yellow in the three-color $Spitzer$ images consisting of the 3.6, 8.0, and 24 $\mu$m bands. Compact \ion{H}{2} regions are visually similar and smaller, which either indicates a less evolved or a more distant \ion{H}{2} region. Since Compact \ion{H}{2} regions are a type of \ion{H}{2} region and comprise only a small sample of the total sample of clumps, we group these clumps with \ion{H}{2} regions for all figures and the rest of the paper.

\begin{deluxetable}{lccc}
\tablecaption{Fraction of Clumps above Regression Fit for Different Evolutionary Stages  \label{tab:percent}}
\tablehead{\colhead{Molecular} & \multicolumn{3}{c}{\underline{Clump Fraction above Regression Fit}} \\
\colhead{Line} &\colhead{\ion{H}{2} Regions} & \colhead{Protostellar} & \colhead{Other}
}
\startdata
HCO$^+$(1--0) & 0.61 $\pm$ 0.05 & 0.33 $\pm$ 0.07 & 0.63 $\pm$ 0.17 \\
HNC(1--0) & 0.63 $\pm$ 0.05 & 0.30 $\pm$ 0.07 & 0.67 $\pm$ 0.19 \\
N$_2$H$^+$(1--0) & 0.60 $\pm$ 0.05 & 0.35 $\pm$ 0.06 & 0.83 $\pm$ 0.15 \\
HCN(1--0) & 0.58 $\pm$ 0.05 & 0.39 $\pm$ 0.07 & 0.86 $\pm$ 0.13 \\
HC$_3$N(10--9) & 0.64 $\pm$ 0.15 & 0.44 $\pm$ 0.12 & N/A \\
C$_2$H(1--0)\tablenotemark{a} & 0.47 $\pm$ 0.09 & 0.46 $\pm$ 0.14 & N/A \\
\enddata
\tablecomments{Uncertainty in the percentages $p$ were calculated via $\sqrt{p(1-p)/n}$ where $n$ is the number of clumps containing a particular evolutionary stage. Percentages marked ``N/A" had insufficient numbers to calculate a percentage and uncertainty.}
\tablenotetext{a}{See Table~\ref{tab:m90} for specific transition.} 
\end{deluxetable}

We indicate the evolutionary classes of each clumps in the scatterplots of Figures \ref{gaosolomon_IRASimfit}, \ref{normalized}, and \ref{quantile}. For the four brightest lines, \hcop, HNC, \nthp, and HCN, there is an apparent trend based on the clump's evolutionary stage; clumps classified as \ion{H}{2} regions tend to be above the least-squares regression fit and clumps classified as Protostellar tend to be below. In Table~\ref{tab:percent} we show the fractions of each evolutionary stage above the fit and show the uncertainty, confirming that this trend is significant.  For the low \lir\ regime, the evolutionary stage of a clump has a clear impact on the relation between \lir\ and \lmol. This is expected because clumps containing \ion{H}{2} regions will have hotter gas and dust than those containing protostellar sources. Continuum flux is more dependent on the dust temperature ($F_d \propto T_d^{4+b}$, where the dust opacity index $b \approx 1-2$) than line flux is on the gas kinetic temperature ($F_g \propto T_k$), and at these high densities, $T_d \approx T_k$. However, this trend becomes less prominent for MALT90 clumps with \liriras~$>~10^{4.5}~L_\odot$. We visualy inspect the $Spitzer$ mid-IR for these particular clumps containing protostellar sources and find that these clumps are particularly bright sources. While early-B stars can also have high luminosities, these clumps could also contain hyper- or ultra-compact \ion{H}{2} regions that are undetectable with $Spitzer$. Moreover, these clumps often have \ion{H}{2} regions associated with the region that are within the IRAS beam. In other words, the Protostellar classification for the more infrared-luminous Protostellar clumps may in fact be \ion{H}{2} regions.



\section{Discussion}\label{discussion}
In order to explain why the \gsr\ extends linearly over several orders of magnitude from galaxies to Galactic clumps, Wu05 proposed that  $\sim$1~pc sized dense clumps are the basic unit of star formation, and these dense clumps with \liriras~$>~10^{4.5}~L_\odot$ have a characteristic ratio \lir/\lhcn\ (where \lhcn~$\propto$~\mdense). They suggested that the observed \lir/\lhcn\ for a galaxy is a summation of large numbers of clumps. At the $\sim$1~pc clump-scale, we question a characteristic value for the ratio \lir/\lhcn\ for the following reasons:
 
\begin{itemize}

\item This interpretation requires approximately 10 times more high-mass star-forming clumps than are currently known in the Milky Way.
\item In many regions and perhaps over an entire galaxy, extended infrared emission can dominate over the infrared emission emitted from clumps, and the regions of extended emission do not follow the \gsr.
\item In the CMZ, the \lir/\lhcn\ ratio is a factor of 4 smaller than that suggested by the \gsr.
\item At the 1~pc clump-scale, there is a significant scatter in the plots for \liriras\ and \lhcn\ (and other molecules); the difference between \lir/\lhcn\ for quantiles $q=0.1$ and $q=0.9$ (the middle 80\% of the data) is approximately two orders of magnitude (Figure~\ref{quantile}).
\end{itemize}

Based on these reasons, we suggest and will assume from hereon that \lir/\lhcn\ is not constant at the clump scale. The \gsr, however, demonstrates that a constant \lir/\lhcn\ exists on the $\sim$10~kpc scales of the entire galaxies. The theoretical framework put forth by \citet{Kruijssen2014} suggests that due to incomplete sampling of independent star-forming regions, star formation relations such as the \gsr\ will no longer hold below some spatial scale. They suggest that the star formation relations will break down for normal spiral galaxies at size-scales of approximately the Toomre length, which is approximately the separation between two spiral arms. If this theoretical framework is correct, then the scatter we find at the clump scale is expected.

Since \lir/\lhcn\ does not appear to be constant at the clump scale, at first glance it may be surprising that the Galactic clumps are scattered amongst the best fit for the \gsr. The entire Galaxy has significant infrared emission from hot, warm, and cold dust \citep{Cox1986}, and each clump is composed of different quantities of each, causing significant scatter for the luminosities of Galactic clumps. Therefore, high-mass clumps may be expected to lie scattered along the \gsr, with the median clump reflecting the galaxy-scale characteristic ratio of \lir/\lhcn.


An understanding of why the \gsr\ is constant at a galaxy-scale requires an understanding of the origin of the infrared and dense gas (as probed by HCN) luminosity. The vast majority of the infrared luminosity comes from reprocessed light generated by high-mass stars, regardless of whether the dust emission comes from regions of compact gas or more diffuse gas heated by the ISRF. All stars form in areas of dense gas, and these areas may dominate the luminosity of HCN. If the ClMF and IMF are both universal at some size-scale, such as the size-scale of a galaxy, integrating the IR and dense gas luminosities of stars and clumps predicted by these mass functions will cause the ratio \lir/\lhcn\ to be constant for the same size-scale. In other words, if the stellar and ClMFs vary from galaxy to galaxy, the ratio \lir/\lhcn\ is not expected to be constant for galaxy-scales. If both the ClMF and IMF are universal at some size-scale, the amount of dense gas from clumps that eventually forms stars must be constant at this size-scale as well, i.e., there is a constant dense gas star formation efficiency (SFE) at this scale. Since gas flows from clumps to cores (size-scale of $\sim$0.1~pc) to stars, a universal ClMF, IMF, and SFE suggests that the core-mass function (CMF) must be universal at the same size-scale. We note that we have not proven that these SFE, IMF, CMF, and ClMF are universal, but suggest that our analysis is consistent with them being universal. Additionally, the universality may only be valid for the types of galaxies observed in GS04, i.e., normal spiral galaxies, LIRGs, and ULIRGs.



The spatial scale at which both the IMF and ClMF are universal has to be sufficiently large in order to sample large populations of high-mass stars and high-mass star-forming clumps. The size-scale must sample areas with cold, warm, and hot dust since all three significantly contribute to a galaxy's total \lir\ \citep{Cox1986}. Moreover, this size-scale needs to sample clumps of multiple stages of evolution since even the evolutionary state of a clump can affect the observed ratio of \lir/\lhcn\  (Section \ref{sec:evol}). This size-scale is certainly much larger than a clump ($\sim$1~pc). Even star formation activities within GMCs (size-scales $\sim$10-100~pc) can vary significantly from cloud to cloud \citep[e.g.,][]{Fukui1999}. Moreover, high-mass stars typically form in spiral arms, so the-size scale where \lir/\lhcn\ is expected to be constant is probably larger than gaps between spiral arms \citep[$\sim$kpc scale, e.g.,][]{Reid2014}. We propose that the ratio \lir/\lhcn\ becomes constant at some scale larger than $\sim$1~kpc because only this size-scale will contain enough clumps to completely sample the ClMF, IMF, and the distribution in their evolutionary states. The idea that the \gsr\ breaks down at some size-scale and depends on the star-forming region's evolutionary stage is consistent with the theoretical framework put forth by \citet{Kruijssen2014}.

The universality of the SFE, IMF, CMF, and ClMF might be questionable due to some scatter in the GS04 plot; the tabulated \lir/\lhcn\ for each galaxy in GS04 can vary by an order of magnitude. This scatter is partially due to the fact that the galaxies are subject to different physical conditions. We suggested in Section~\ref{largecmz} that the physical conditions toward a galaxy's center may be a source of this scatter. In general, the SFE, IMF, CMF, and ClMF may not be universal toward Galactic nuclei where the environments may be significantly different (e.g., different levels of turbulence and whether the galaxy contains an active galactic nuclei). Galaxies can also be at different evolutionary stages or subject to different conditions (e.g., starbursts or mergers). \citet{Gracia2008} found that \lir/\lhcn\ may be a factor of 2--3 higher in LIRGs and ULIRGs than in normal galaxies. This difference could be due to the fact that emission from high-density tracers can be self-absorbed toward the nuclei (few 100 pc scale) of LIRGs and ULIRGs \citep{Aalto2015}, and these nuclei contain a large fraction of these galaxies' IR luminosity.  \citet{Gao2007} found that in early universe ($2 <  z < 6.5$) emission-line galaxies (EMGs) are significantly offset from the \gsr, with a large excess of \lir\ per unit of \lhcn. The galaxies in GS04 all have $z<0.1$ (and typically have $z<0.01$), suggesting the star-forming properties of high-redshift galaxies may differ significantly from those of low-redshift galaxies. 
EMGs typically have lower metallicities\footnote{In order to calculate metallicity in units of $Z_\odot$, we assume 12 + (O/H)$_\odot$ = 8.66 \citep{Asplund2004} in this paper.} \citep[0.07--0.7~$Z_\odot$ for $0.6 < z < 2.4$,][]{Xia2012} which may account for the deficit of \lhcn\ compared to GS04 galaxies. Metallicities of LIRGs and ULIRGs can also vary significantly, from $\sim$0.3 to 2.8~$Z_\odot$, with typical values $\sim$1.0--1.25~$Z_\odot$ \citep{Rupke2008}. While metallicity differences are not expected to affect the IMF \citep[][and references therein]{Bastian2010}, they certainly affect the abundances of high-density tracers like HCN due to increased photoionization of the molecules from a stronger radiation field \citep[e.g.,][]{Bolatto1999} and the availability of metals. On the other hand, large grains, the primary emitters for IRAS bands, are more resilient to destruction for a given radiation field \citep[e.g.,][]{Stephens2014a}. In short, just as clumps are subject to different conditions within the Milky Way, external galaxies are also subject to different conditions which could be the source of scatter for \lir/\lhcn\ in GS04.





Since clumps do not appear to have a specific \lhcn/\lir\ ratio as proposed by Wu05 and Wu10, a different explanation is needed for the reason that \lir\ versus \lcooz\ follows a different power-law relation than \lir\ versus \lhcnoz. \citet{Narayanan2008} considered a galaxy that follows the Schmidt law \citep{Schmidt1959}, i.e., the star formation rate (SFR, typically probed by \lir) depends on the mean molecular gas mass density $\rho$ according to

\begin{equation}\label{schmidt}
\mbox{SFR} \propto \rho^N .
\end{equation}
In the \citet{Narayanan2008} simulations, which includes 3D non-local thermodynamic equilibrium radiative transfer calculations, the luminosity of the molecules (at least those considered in the simulations, i.e., CO and HCN), follow the power-law

\begin{equation}
L_{\rm{molecule}} \propto \rho^\beta,
\end{equation}
where $\rho$ is the mean molecular gas mass density in a grid cell of clouds along the line of sight. Given these two equations, the relation between SFR (as probed by \lir) and \lmol\ can be understood as

\begin{equation}
L_{\rm{IR}} \propto \mbox{SFR} \propto L_{\rm{molecule}}^\alpha,
\end{equation}

\noindent where $\alpha = N/\beta$. Equation~\ref{schmidt} is commonly assumed to behave according to the Schmidt index $N=1.4$. Therefore, different values of $\beta$ for different molecules reflect the power-law index $\alpha$ observed across an entire galaxy. \citet{Narayanan2008} suggested that the value of $\beta$ depends on the amount of diffuse gas emitting subthermally by molecules resonantly scattering photons emitted from dense regions. Typically \cooz\ gas is above the critical density, causing there to be an insignificant emission from subthermal \cooz\ gas. Therefore, the luminosity of CO will be directly proportional to the density, i.e., $\beta=1$, and the SFR--line luminosity relation behaves according to the Schmidt index, i.e., $\alpha = N = 1.4$. However, high-density tracers such as \hcnoz\ have critical densities which are above the average number densities of a typical galaxy. Therefore, a large quantity of \hcnoz\ emitting gas will be diffuse, emitting subthermally. The simulations from \citet{Narayanan2008} suggested that emission from thermally emitting dense clumps can be absorbed by diffuse gas and re-emitted subthermally. The diffuse \hcnoz\ gas will have a higher luminosity per unit density since the photons originate from thermal emission of dense regions. Therefore, the \lmol -- $\rho$ relation will be superlinear, i.e., $\beta > 1$, causing $\alpha$ to be less than the Schmidt index. HCN appears to have $\beta = N$ causing $\alpha = 1$. Other lines can have a variety of values of $\beta$ (larger than 1), allowing for the possibility of $\alpha$ to have a value of less than 1. $\alpha$ values of sub-unity have been found for galaxies using higher-$J$ transitions \citep{Baan2008,Gracia2008, Bussmann2008, Juneau2009} in accordance with the predictions of \citet{Narayanan2008}. Nevertheless, \citet{ZhangZY2014} observed even higher-$J$ transitions with APEX, specifically HCN(4--3), HCO$^+$(4--3), and CS(7--6), and found values of $\alpha$ near unity, disagreeing with \citet{Narayanan2008}. 


Our analysis in Section~\ref{subthermal} suggests significant subthermal emission of high-density tracers within the Milky Way and other galaxies. This emission may cause $\beta$ to be larger than 1. Therefore, our observations are consistent with the simulations of \citet{Narayanan2008}. Nevertheless, we are unable to confirm that $\beta = N$ and that a significant amount of diffuse HCN emission arises from resonant scattering of photons originally produced in dense clumps.


\section{Summary}\label{summary}
We use molecular line data from clumps ($\sim$1 pc scale) from the MALT90 survey, and, along with other publications and ancillary data, we investigate the \gsr\ for HCN and the \lir\ versus \lmol\ relation for 11 other molecular lines. We reach the following conclusions:\\
1) We reject the Wu05 idea that clumps have a specific \lir/\lhcn\ ratio (with small scatter) for clumps $>$10$^{4.5}~L_\odot$. The ratio \lir/\lhcn\ is probably not constant at the clump-scale, but is constant at some larger scale. We propose that the scale in which \lir/\lhcn\ is expected to be constant is $\gtrsim$1~kpc.\\
2) High-mass star-forming clumps likely account for only $\sim$10\% of an entire galaxy's \liriras\ and \lhcn. Much of the IR emission comes from dust heated by the ISRF. The dominant source of a galaxy's \lhcn\ is uncertain, but we suggest that low-mass star-forming clumps or subthermal emission may dominate the HCN emission.
3) Our analysis is consistent with the models set forth by \citet{KrumholzThompson2007} and \citet{Narayanan2008} that suggest that the relation \lir~$\propto$~\lmol$^\alpha$ depends on how the molecular line's critical density compares with the median density of star-forming clouds in a galaxy. For lines with critical density significantly above the median density of a galaxy, a significant amount of photons are redistributed from thermal gas to subthermal gas, causing $\alpha$ to be smaller than the Schmidt index of $N=1.4$.\\
4) The fact that \lir/\lhcn\ is only expected to have the same characteristic value of galaxies at large scales is consistent with the idea that spiral galaxies have a universal star formation efficiency, initial mass function, core mass function, and clump mass function. While the universality can be questioned at small scales, they are likely universal at some large scale, which we suggest to be $\gtrsim$1~kpc.\\
5) The Central Molecular Zone (CMZ) adds significant luminosity to the entire Milky Way's \lir\ and \lhcn, but the CMZ does not lie on the best fit line of the \gsr. This is likely due to the fact that the CMZ is subject to different conditions, e.g., extremely high turbulence, as compared to the rest of the galaxy. Differences in the contribution of other galaxies' CMZ to the global HCN and IR luminosities could be a source of scatter for the \gsr.\\
6) We investigate the Galactic clump relation between \lir\ and \lmol\ for molecules other than HCN. A positive correlation exists between \lir\ and \lmol\ for each molecule, and we find no sudden drop in the ratio of \lir/\lmol\ for \lir\ $<~10^{4.5}~L_\odot$. We find that the \lir/\lhcn\ ratios are the most similar to \lir/\lhcop\ ratios.\\
7) The evolutionary stage of a clump, particularly for clumps with lower infrared luminosities, helps to predict whether a clump will have a higher or lower \lir/\lhcn\ ratio than expected from the \gsr. Specifically, protostellar sources are more likely to have a lower \lir/\lhcn\ ratio than \ion{H}{2} regions.

In short, this paper finds that the \gsr\ is not likely explained as a summation of high-mass star-forming clumps. The relation could be explained by a universal star formation efficiency, initial mass function, core mass function, and clump mass function. The size-scale at which \lir/\lhcn\ becomes a constant value must be much larger than the clump-scale. Future observations that map entire galaxies with high-density tracers at sub-kpc resolution will determine this size-scale. 


\acknowledgments
We thank Mark Reid for pointing out the large amount of high-mass clumps that would be required to account for the global Galactic HCN and IR luminosities. This work was supported by NASA grant NNX12AE42G and NSF grant AST-1211844. A.E.G. acknowledges support from FONDECYT grant 3150570. This research made use of APLpy, an open-source plotting package for Python hosted at http://aplpy.github.com. 

\bibliography{stephens_bib}

\appendix
\section{Table of Fluxes for Each Molecular Line}
\clearpage
\LongTables
\begin{landscape}
\begin{deluxetable}{lcccccccccccccccc}
\tabletypesize{\scriptsize}
\tablewidth{0pt}
\setlength{\tabcolsep}{1pt}
\tablecaption{IR and Molecular Fluxes \label{tab:main}}
\tablehead{\colhead{ATLASGAL\tablenotemark{a}} & \colhead{Mid-IR\tablenotemark{b}} & \colhead{D\tablenotemark{c}} & \colhead{\firiras} & \colhead{\fhcop} & \colhead{\fhnc}  & \colhead{\fnthp}  & \colhead{\fhcn}  & \colhead{\fhtcop}  & \colhead{\fhntc}  & \colhead{\ftcs}  & \colhead{\fhncofzf}  & \colhead{\fchtcn}  & \colhead{\fsio}  & \colhead{\fhctn} & \colhead{\fcth}   \\
\colhead{Source} & \colhead{Classification} & \colhead{(kpc)} & \colhead{log Scale ($L_\odot\,$kpc$^{-2}$)} & \multicolumn{12}{c}{log Scale (\Kkmspcsq $\,$kpc$^{-2}$)} 
}
\startdata
AGAL000.166-00.446\_S & HII Region & ... & 2.58 & ... & ... & 0.54 & ... & ... & ... & ... & ... & ... & ... & ... & ... \\
AGAL000.281-00.482\_S & HII Region & ... & 3.43 & 0.96 & 0.75 & 0.60 & 0.95 & ... & ... & ... & ... & ... & ... & ... & ... \\
AGAL000.316-00.201\_S & HII Region & ... & 3.91 & 1.13 & 0.82 & 0.62 & 1.10 & ... & ... & ... & ... & ... & ... & 0.01 & 0.47 \\
AGAL000.496+00.187\_S & HII Region & ... & 3.39 & ... & ... & 0.54 & 0.17 & ... & ... & ... & ... & ... & ... & ... & ... \\
AGAL000.526+00.182\_S & HII Region & ... & 3.41 & 0.10 & 0.12 & 0.22 & 0.24 & ... & ... & ... & ... & ... & ... & ... & 0.25 \\
AGAL000.631+00.604\_S & PDR & ... & 2.79 & ... & ... & -0.22 & ... & ... & ... & ... & ... & ... & ... & ... & ... \\
AGAL000.666-00.034\_A & Protostellar & ... & 4.75 & 1.62 & 1.38 & ... & 1.80 & 0.22 & ... & ... & 0.75 & 0.72 & 0.63 & ... & 0.27 \\
AGAL000.666-00.034\_B & Protostellar & ... & 4.75 & ... & ... & ... & ... & 0.70 & 0.69 & 0.61 & ... & 1.10 & ... & 1.44 & 1.08 \\
\enddata
\tablecomments{This table shows a list of all MALT90 sources that match with an IRAS source.}
\tablenotetext{a}{ATLASGAL clumps have either $\_$S, $\_$A, or $\_$B after there names, as per J. Rathborne et al. (in preparation). Those with the nomenclature $\_$S are not likely to have multiple clumps along the line of sight. Those with $\_$A and $\_$B are likely to have confusion along the line of sight (i.e., they have two velocity components separated by $>$15~\kms). For all numbers, fits, and figures within the bulk of the paper, $\_$A and $\_$B components are not used since the IRAS continuum contributed to each the $\_$A and $\_$B components cannot be deduced. We report the fluxes here for completeness. AGAL344.582-00.024\_S and AGAL346.418+00.279\_S luminosities were not used in this paper because the distances are unrealistically close for \ion{H}{2} regions.
} 
\tablenotetext{b}{Classification based on $Spitzer$ three-color images (see Section~\ref{sec:evol}).} 
\tablenotetext{c}{Distances are kinematic distances form J. Whitaker et al. (in preparation).} 
\end{deluxetable}
\clearpage
\end{landscape}

\end{document}